# Annual Reviews
# of Astronomy and Astrophysics

# The Pluto System After New Horizons

S. A. Stern
Southwest Research Institute
alan@boulder.swri.edu
303-546-9670

W. M. Grundy
Lowell Observatory
grundy@lowell.edu

Wm. B. McKinnon
Washington University in St. Louis
mckinnon@wustl.edu

H. A. Weaver
Johns Hopkins Applied Physics Laboratory
Hal.Weaver@jhuapl.edu

and

L. A. Young
Southwest Research Institute
leslie@boulder.swri.edu

Corresponding Author: S. Alan Stern,
Southwest Research Institute, Suite 300,
1050 Walnut Street, Boulder Colorado,
Email: astern@swri.edu.
Phone: 303-324-5269

**Abstract**

The discovery of Pluto in 1930 presaged the discoveries of both the Kuiper Belt and ice dwarf planets—the third class of planets in our solar system. From the 1970s to the 1990s, numerous fascinating attributes of this binary planet were discovered, including multiple surface volatile species, the presence of its largest satellite Charon, and its atmosphere. These attributes, and the 1990s discovery of the Kuiper Belt and Pluto's cohort of small Kuiper Belt planets motivated the spacecraft exploration of Pluto. That mission, called New Horizons (NH), revolutionized our knowledge of Pluto and its system of moons in mid-2015. Beyond providing rich geological, compositional, and atmospheric datasets, NH demonstrated that Pluto has been surprisingly geologically and climatologically active throughout the past 4+ Gyr, and that the planet exhibits a surprisingly complex range of atmospheric phenomenology and geological expressions that rival Mars in their richness.

# 1. Introduction

## 1.1. The Pluto System Before NH

Pluto was discovered in 1930 by Clyde W. Tombaugh, following a quarter-century long search primarily prompted by Percival Lowell (see Tombaugh & Moore 1980 for a historical account). Owing to its faintness and small angular diameter as seen from Earth, little could be gleaned about the planet using the technology of the mid-20th century. Indeed, except for its orbital characteristics ($a$=49.6 AU; $e$=0.25, $i$=17 deg), and extremely crude size estimates, nothing significant was learned about Pluto for decades after its discovery. As reviewed in Stern (1992), even through 1975 only its lightcurve (yielding a rotational period of 6.3871 Earth days and a surprisingly-high ~25% amplitude), its color (B-V=0.85), its high axial obliquity (121 deg), and its heliocentric dynamical location in a 3:2 mean motion resonance with Neptune were determined.

However, the late 1970s yielded two key discoveries that can be considered the beginning of the modern era of groundbased Pluto system observations. First was the discovery of the volatile ice $CH_4$ on Pluto's surface (Cruikshank et al. 1976). This in turn initiated speculation that Pluto might harbor an atmosphere, and indicated a high surface albedo, meaning Pluto's size was therefore likely smaller than had been expected. Second, in 1978, an astrometry program being carried out by the U.S. Naval Observatory revealed the presence of a large satellite in a synchronous, equatorial, circular orbit above Pluto (Christy & Harrington 1978). This satellite was later named Charon (usually pronounced 'Sharon," for Christy's wife); see Stern (1992) for more on these and other early findings.

The late 1970s and the 1980s produced further advances, most made possible by the prediction (Andersson 1978) and the first detection (Binzel et al. 1985) of a fortuitous multi-year long series of mutual Pluto-Charon occultation events observable from Earth. These mutual events (see Binzel 1989 and Stern 1992) yielded many discoveries including: (i) primitive maps of Pluto's Charon-facing hemisphere; (ii) proof that the $CH_4$ in the combined spectrum of Pluto and Charon is on Pluto's surface (not Charon's); (iii) the discovery of $H_2O$-ice covering Charon's surface; (iv) the first relatively accurate radii of Pluto and Charon (clearly establishing the pair as a binary planet); and (v) the subsequent discovery of an unexpectedly high ~2 g cm$^{-3}$ average density for Pluto-Charon—somewhat surprisingly indicating primarily *rocky* (vs. icy) bulk compositions. Pluto's atmosphere, having been searched for but not detected spectroscopically and having been the subject of numerous theoretical considerations (e.g., Trafton 1980; Trafton & Stern 1983; Stern et al. 1988) was first definitely detected in a 1988 stellar occultation (Elliot et al. 1989; see also Hubbard et al. 1990). This revealed Pluto's surface pressure to be in the ~1-10 microbar domain and its atmosphere to be highly distended with evidence for either low altitude hazes or a large vertical thermal gradient, or both. The 1980s also saw the case first made for the giant impact formation hypothesis of the Pluto-Charon binary (McKinnon 1989).

The 1990s and 2000s produced many additional discoveries, most made possible by advances in detector technologies and the advent of both larger groundbased telescopes and the Hubble Space Telescope (HST). Those discoveries included (i) the detection of the volatile ices $N_2$ and CO on Pluto's surface (Owen et al. 1993); (ii) the first spectroscopic detections of Pluto's atmosphere (Young et al. 1992); (iii) the first measurements of the separate (and somewhat different) densities of Pluto and Charon (e.g., Null et al. 1993; Foust et al. 1997; Olkin et al. 2003); (iv) the first direct imaging of features on Pluto's surface (Stern et al. 1997a); (v) the first measurement of Charon's surprisingly low-amplitude (8%) lightcurve (Buie et al. 2002); (vi) the discovery of Pluto's first two small satellites (Weaver et al. 2006; Stern et al. 2006); (vii) the detection of changing atmospheric pressure and vertical structure on Pluto (Elliot et al. 2006; Sicardy et al. 2006); (viii) the detection of temporal change on Pluto's surface (e.g., Buie et al. 2010b); and (ix) reports of $NH_3$-bearing compounds on Charon's surface (e.g., Cook et al. 2007).

## 1.2. Pluto's Context in the Solar System

Pluto's small size and unusual orbit compared to the giant planets prompted various speculations about its origin and provenance from the 1930s to the 1990s. Notable papers included Lyttleton (1936), which speculated on the now disproven idea that Pluto was an escaped satellite of Neptune, and papers by Edgeworth (1943, 1949), and most famously Kuiper (1951) suggesting Pluto was the harbinger of a large "trans-Neptunian" population of comets and larger bodies. Strong dynamical evidence for that population, based on the orbital inclination distribution of short period comets, was later found by Duncan et al. (1987). This led to a number of telescopic searches for that cohort population, which was finally detected by the discovery of the first Kuiper Belt (KB) Object (Jewitt & Luu 1993). Subsequently, a vast population of other KB Objects (KBOs) was discovered (see papers in Barucci et al. 2008) in a disk- or torus-like structure and a more distant, extended, scattered halo that together are now widely recognized as our planetary system's third zone, beyond the inner zone of the inner planets and the middle zone of the giant planets.

This structure, known as the KB (KB), has provided many new constraints on planet formation in the outer solar system and gave the first definitive evidence for the giant planet migration (Malhotra 1993). The KB's population is comprised of comets (~1-20 km diameter bodies), planetesimals (50-300 km in diameter), and small planets (300-2400 km in diameter) of which Pluto is the largest. Additionally, the KB was revealed to have a complex dynamical substructure, a variety of color populations, and a significant fraction of bodies with satellites. Among the discovered dynamical classes of KBOs are "Plutinos"—other bodies also in the 3:2 Neptune Mean Motion Resonance (MMR), and bodies in other low order multiple Neptunian MMRs, both of which have migrated outward with the giant planets; also discovered was both a dynamically cold KB population called the Cold Classical KBOs which apparently formed more or less in situ, and a dynamically hot population of KBOs that have apparently been scattered outward by one or more violent dynamical rearrangements involving the giant planets early in the history of the solar system (e.g., Levison et al. 2008).

Perhaps most importantly, the KB also yielded detections of a cohort population of small (1000-to-2400 km diameter) planets like Pluto, which were not part of Kuiper's original hypothesis, but which had been later predicted based on forensic clues around the outer solar system (Stern 1991). These discoveries demonstrated the existence of a third class of planets in our solar system—called the "dwarf planets" (coined by Stern (1991)). As we now know, there are of order ten 1000-km diameter or larger bodies orbiting in the KB of which Pluto is now definitively established to be the largest (Stern et al. 2015). Many more are likely to orbit beyond the KB in the Oort Cloud (OC). This cohort population shows a wide heterogeneity (as do the terrestrial planets) in terms of their surface compositions, bulk densities, satellite populations, and other attributes (again see papers in Barucci et al. (2008)).

### 1.3. Spacecraft Reconnaissance of the Pluto System by NH

The first serious discussion of spacecraft reconnaissance of the Pluto system began at the end of the 1980s and continued through a wide variety of NASA mission study concepts and advisory committee reports spanning the 1990s (for more on this history, see Stern 1993, Stern 2008, Weaver & Stern 2008, Neufield 2016; see also Grinspoon & Stern 2018). All of these studies involved flyby reconnaissance spacecraft with various capabilities, some even involving two spacecraft flying by Pluto like the early mission pairs sent to reconnoiter Venus, Mars, and the giant planets.

The scientific case for studying the Pluto system with a dedicated spacecraft mission or missions was based primarily upon (i) Pluto's many fascinating surface, satellite, and atmospheric attributes and (ii) the time criticality imposed by Pluto's motion outward from the Sun which prompted a concern that its atmosphere might cool and pressure collapse before a reconnaissance spacecraft could arrive to study it. Numerous NASA and NRC advisory committees (e.g., Lunine et al. 1996) endorsed the flyby reconnaissance of Pluto in the 1990s, but secure funding for NASA to carry out this project only resulted when the first Planetary Decadal Survey ranked this exploration as a top priority (Belton et al. 2003). That high priority was fundamentally motivated not just by the fascinating attributes of the Pluto system, but also by the discovery of the KB and its context and importance to understanding the solar system's formation. Of further importance was the new perspective that Pluto was seen as the first discovered, the brightest, and the best studied of the new class of objects beyond Neptune called the dwarf planets of the KB.

The NH mission itself resulted from a competition NASA held in 2001 for Pluto flyby missions. The highest priority objectives that NASA set out for this mission were to map the surfaces of Pluto and Charon, to map their compositions, and to determine the composition, vertical pressure-temperature structure, and escape rate of Pluto's atmosphere. Secondary and tertiary objectives included higher resolution geological and compositional studies of selected terrains, stereo mapping of terrain elevations, searches for trace species in Pluto's atmosphere, surface temperature measurements, the refinement of bulk parameters for Pluto and Charon, and searches for additional satellites and rings (Stern 2008; Weaver & Stern 2008; Young et al. 2008).

NH, a single spacecraft flyby reconnaissance mission, was launched by an Atlas V 551 rocket on 19 January 2006 onto a direct ascent trajectory to Jupiter, where it executed a Jupiter Gravity Assist on 28 February 2007 to target a Pluto flyby that occurred on 14 July 2015. Its study of the Pluto system and its environment spanned a six-month period from January through July of 2015.

The NH spacecraft carries a payload of seven advanced instruments (Weaver et al. 2008; Cheng et al. 2008; Reuter et al. 2008; Stern et al. 2008; Tyler et al. 2008; McNutt et al. 2008; McComas et al. 2008; Horanyi et al. 2008), which include the following key capabilities: medium and high-resolution panchromatic visible wavelength mapping, medium resolution visible wavelength color mapping; IR surface composition mapping; stereo imaging for terrain height mapping; ultraviolet spectroscopy for atmospheric composition and vertical structure studies; plasma spectrometers to measure the escape rate and the composition of ionized gases escaping from Pluto's atmosphere, as well as the interaction of the atmosphere with the solar wind; radio science to measure the brightness temperature of Pluto's surface, to determine the vertical temperature-pressure profile of Pluto's lower atmosphere, and to make bistatic radar measurements at 4.2 cm wavelengths; and a dust detector to search for particulates in orbit around Pluto. More detail on the instrument payload and its scientific objectives can be found in (Weaver & Stern 2008).

The flyby of the Pluto system by NH was completely successful (Stern et al. 2015), meeting and in many cases exceeding the objectives set out for it by NASA and the National Academy of Sciences (Lunine et al. 1996, Belton et al. 2003). In addition to studying Pluto and Charon as originally intended, NH also studied all four of Pluto's small satellites (all discovered after the spacecraft was built). In what follows we review the major findings of this exploration and their implications in greater detail.

### 1.4. Nomenclature

Most place names on Pluto and its satellites are as of this writing informal, awaiting formal approval.

Herein we refer to the planets of the KB as dwarf planets (DPs). As planetary scientists we prefer the Geophysical Planet Definition (GPD; e.g., Runyon et al. 2017) to the 2006 IAU definition, and we, like others, consider DPs full-fledged planets owing to the similarity of processes and attributes to larger planets. The discovery of geological and atmospheric complexity and activity made by NH validates this rationale. We predict that the KB's scattered disk and the Oort Cloud (OC) contain many more DPs than have been discovered in the KB to date (Stern 1991).

We also consider satellites of planets large enough to satisfy the GPD to be planets that orbit other planets. Hence, both Pluto and Charon are referred to here as planets; and because the system barycenter is external to both objects in the binary in free space between them, the system is also referred to here (as it often has been elsewhere) as a binary planet.

## 2. Pluto

Pluto is the largest known planet in the KB and accounts for 89% of the mass in a system comprised of Pluto and Charon, plus four much smaller, outer satellites. Pluto is now known to have been geologically active through solar system history, continuing to the present day. It has a differentiated interior with volatile veneer above an icy mantle overlying a rocky core; it is massive enough to retain an atmosphere. Pluto's interior, surface, and atmosphere interact in diverse ways over a broad range of timescales, producing a complex, dynamic world.

### 2.1. Geology and Composition

Pluto's surface exhibits spectacular geological diversity. Some landforms are familiar from spacecraft exploration elsewhere, such as impact craters and fault scarps, but processes enabled by Pluto's inventory of volatile materials produce more exotic features as well. Ices of $N_2$, CO, and $CH_4$ are all present on Pluto's surface at temperatures in the 30 to 60 K range. These ices can be remotely identified and mapped via reflectance spectroscopy of their characteristic near-infrared absorption bands (e.g., Owen et al. 1993, 2003). Despite the low energy inputs available to mobilize them (~1 W m$^{-2}$ from sunlight and ~0.003 W m$^{-2}$ from decay of radionuclides in Pluto's interior) these volatile ices evolve on seasonal and longer "mega-seasonal" timescales, via sublimation and condensation as well as by glacial flow and likely cryovolcanism. Underlying Pluto's surface volatiles is an $H_2O$-ice dominated mantle above an inferred rocky core (§2.3).

Here we first consider Pluto's more conventional geology (Figure 1), involving impact and tectonic modification of the relatively inert, $H_2O$ ice dominated mantle "bedrock", and then turn to geology enabled by its more volatile materials.

Impact craters are abundant on Pluto. Over 5000 possible/probable craters have been mapped, almost all on the encounter hemisphere, where NH obtained its highest resolution images (Moore et al. 2016; Robbins et al. 2017; Singer et al. 2017). Their distribution is nonuniform: crater-free regions could be as young as a few tens of Myr or less, whereas parts of Venera, Voyager, and Viking Terrae, for example, have retained craters over ~4 Gyr timescales. Small craters exhibit simple bowl-shaped morphology, while some larger craters show central uplifts and other more complex morphologies, as have been observed on other bodies (e.g., Melosh 1989). However, ejecta blankets are difficult to discern, likely due to seasonal volatile ice deposition. In some regions, rims of craters exhibit strong spectral signatures of $CH_4$ ice, likely due to its seasonal deposition on high-standing terrain (Grundy et al. 2016a; Schmitt et al. 2017). Other craters show smooth deposits of dark, reddish material on their floors. Such modification complicates the use of craters as windows into Pluto's interior, but some craters show dark, horizontal banding in their interior walls indicative of subsurface strata. Many craters are eroded, sometimes with a radial, fluted pattern.

A number of large troughs and scarps extend hundreds of km across the planet's encounter hemisphere, especially in Viking Terra and Cthulhu Macula, with up to a few km vertical displacement. Some show evidence of subsurface strata in their exposed faces. Their morphology is consistent with extensional tectonics (Moore et al. 2016). Some scarps appear to be oriented radial to Sputnik Planitia (Keane et al. 2016), a ~1000 km wide oval depression hypothesized to be an ancient impact basin (e.g., Stern et al. 2015; Moore et al. 2016). The radial fractures could result from its formation and/or post-formation stress/strain evolution (Keane et al. 2016; Hamilton et al. 2016; Nimmo et al. 2016). Other tectonic features have orientations that may be consistent with the reorientation of Pluto relative to its spin and tidal axes (see §2.3). Multiple trough/scarp orientations and a range of degradation states suggest a long period and/or multiple episodes of tectonic activity (Moore et al. 2016). Sleipnir and Mwindo Fossae, in the far east of the encounter hemisphere, is a system of extensional fractures radiating from a common locus (Moore et al. 2016; Keane et al. 2016) that resembles systems on Venus, and which appears to be unique in the outer solar system. It may indicate localized, focused volumetric changes in the subsurface.

Sputnik Planitia (SP) hosts a vast glacial deposit of volatile ices (Stern et al. 2015). $N_2$, CO, and $CH_4$ ices are all detected spectroscopically there, though $N_2$ is the primary constituent (Protopapa et al. 2017). These ices likely accumulated in Sputnik soon after formation of its basin (Bertrand & Forget 2016; Hamilton et al. 2016). No unambiguous impact craters are seen anywhere across Sputnik's ~1,100,000 km$^2$ extent, implying a very young (<10-100 Myr) surface age and recent or ongoing resurfacing (Moore et al. 2016).

Much of SP's surface is marked by a polygonal or cellular network of low ridges, troughs, and lineations (Figure 2; White et al. 2017), interpreted as the boundaries of convection cells several tens of km across (Moore et al. 2016; McKinnon et al. 2016; Trowbridge et al. 2016). Models indicate that if this volatile ice is thicker than ~500 m, radiogenic heat from Pluto's interior is sufficient to drive convection, thanks to the low viscosity and low thermal conductivity of $N_2$ ice. Buoyant plumes of warmer ice rise in the centers of the cells while cooler ice sinks along their peripheries, with an overturning timescale of $10^5$ to $10^6$ years (McKinnon et al. 2016). There is, however, considerable uncertainty about the temperature-dependent rheological properties of $N_2$ ice when it is mixed with $CH_4$, CO, and other plausible contaminants.

On a finer scale, Sputnik's surface is patterned with round to elongated ripples, dimples, and pits a few hundred meters across (White et al. 2017). They vary gradually in character and scale with location. In northwest Sputnik, Telfer et al. (2017) map some of these features as potential dunes, suggesting that they could be composed of wind-blown $CH_4$ ice grains. The morphology shifts to more pit-like forms farther south and east that may be produced by sunlight-driven sublimation and condensation amplifying topographic variations, moderated by glacial relaxation (Moore et al. 2016; Buhler & Ingersoll 2017; Howard et al. 2017a). The density and scale of pitting varies from the centers to margins of the broader cells, consistent with pit formation and evolution timescales comparable to the estimated convective overturning timescale, and also influenced by regional differences in convective vigor and insolation. Elsewhere on Pluto, similar-looking pits can be seen in smaller, low-lying deposits associated with crater and valley floors, where volatile ices also accumulate.

The eastern margin of Sputnik Planitia hosts more familiar-looking glacial morphology (Figure 2). Valley glaciers appear to deliver volatile ice from eastern Tombaugh Regio into Sputnik (Moore et al. 2016; Howard et al. 2017a). Flow lines resembling moraine trails on terrestrial valley glaciers are consistent with down-slope flow through glacially eroded channels (Howard et al. 2017a; White et al. 2017). The flow patterns can be traced 100-200 km into eastern Sputnik Planitia. Models indicate that deposits of $N_2$ ice of sufficient thickness can flow under Pluto thermal conditions, though as noted above, rheological uncertainties remain (see Umurhan et al. 2017).

Small chains of hills extend into Sputnik from the mouths of the valley glaciers that are apparently transported by glacial flow (McKinnon et al. 2016). $H_2O$ ice (~0.9 g cm$^{-3}$) is less dense than the $N_2$ ice (~1 g cm$^{-3}$) that dominates Sputnik and could thus be supported buoyantly. $CH_4$ ice (~0.5 g cm$^{-3}$) is considerably less dense and would float much higher. The hills seen in eastern Sputnik are unresolved by NH' infrared spectrometer, so their composition is uncertain, but much larger mountains along

Sputnik's western flank show spectral features of both $H_2O$ and $CH_4$ ices. It is therefore thought that the hills in Sputnik are "icebergs" of $H_2O$ and/or $CH_4$ floating on the $N_2$ glacier.

The Tenzing, Hillary, Baré, Zheng He, and al-Idrisi mountain ranges punctuate western Sputnik Planitia; some could be much larger versions of the floating hills (Stern et al. 2015; Moore et al. 2016). Resembling rafted and rotated blocks reaching tens of km across and with summits rising to ~5 km above SP, they could be fragments of material broken from the uplands to the west of Sputnik. If their compositions are dominated by rigid $H_2O$ ice, as suggested by their height and steepness, the small $H_2O$–$N_2$ density contrast implies enormous submerged keels, possibly inconsistent with the close proximity of adjacent peaks. An alternate explanation could be that the ice sheet has retreated, consistent with expectation from climate models (Forget et al. 2017), revealing previously submerged roots. Another alternative is that the block compositions could be $CH_4$-rich and thus float higher, due to the lower density of $CH_4$ ice. Both $H_2O$ and $CH_4$ ice absorptions are seen in spectra of the mountains, though it is not clear if these signatures are indicative of bulk composition or just surface veneers.

Diverse channels and valley networks dissect Pluto's encounter hemisphere, as do the glacial valleys draining into eastern SP and the fluted erosion of crater rims we earlier noted (Howard et al. 2017a). These include dendritic channels in Pioneer Terra (Figure 3a) and fretted terrain in western Venera Terra. The flat-floored, sinuous Kupe Vallis resembles an inactive version of the valley glaciers of eastern Tombaugh Regio. In Lowell Regio, Ivanov Vallis extends northward for a few hundred km from a breached crater rim. Such channels could potentially be related to glacial flow of volatile ices during past climatic epochs (Howard et al. 2017a). An alternative to this scenario involves melting at the base of thick glacial deposits and sub-glacial fluid erosion. For reasonable sub-surface temperature gradients, pure $N_2$ would reach its 63 K melting point just a couple of kilometers below Pluto's surface. Mixtures of ices are more probable, and freezing point suppression can occur in some of them, calling for further laboratory work to explore this.

"Washboard" terrain to the northwest of Sputnik Planitia consists of parallel sets of north-south trending low ridges and troughs spaced about 1 km crest to crest, with underlying terrain features remaining visible (Moore et al. 2016). Its origin is uncertain, but may be related to sublimation, aeolian, or glacial processes (Howard et al. 2017a).

Various kilometer-scale pits occur in eastern Tombaugh Regio and also farther north in Hayabusa and Pioneer Terrae (Figure 3b). Clusters of evenly spaced, similarly sized pits tend to form aligned distributions, possibly governed by a local tectonic fabric. Some are especially deep, with symmetric, conical interior profiles, possibly at the angle of repose. Some have elevated rims. Larger, irregularly shaped, flat-floored depressions tens of km across and up to ~3 km deep occur in Pioneer Terra and Lowell Regio. Potential pit-forming processes include collapse into the subsurface (perhaps related to subsurface melting or sublimation loss) as well as eruptive ejection of material, although apart from the raised rims, ejecta deposits are not obvious (Howard et al. 2017b). In many regions where large pits are seen, there is an appearance of hundreds of meters to a kilometer of mantling that smooths the landscape, with the pits being excavated into that material and possibly growing via scarp retreat. The scarp of Piri Rupes appears to have retreated, exposing an $H_2O$-rich substrate in Piri Planitia (Moore et al. 2017b). The scarp and surrounding uplands show spectral signatures of $CH_4$ ice. Likewise, highlands in Pioneer Terra feature $CH_4$, with $H_2O$ on the floors of some pits—perhaps exposed "bedrock"—though other pits are floored with the more volatile ices $N_2$ and CO that preferentially accumulate in topographic lows (Howard et al. 2017b).

Tartarus Dorsae is an especially striking landscape, described as "bladed terrain" (Figure 3c). Narrow, north-south aligned ridges perch atop broader swells (Moore et al. 2016, 2017a). The blades range from a few hundred meters to nearly a km in height, and are spaced several km apart. Impact craters are scarce in this region, suggesting a relatively recent origin. Spectroscopic observations show abundant $CH_4$ ice (Grundy et al. 2016a; Schmitt et al. 2017). Elsewhere on the encounter hemisphere, low latitude $CH_4$ ice deposits are associated with crater rims and other high-standing regions, suggesting that the blades could grow via preferential deposition of $CH_4$ at high altitude, with their orientations influenced by wind-flow and/or sunlight. Another idea regarding these landforms involves sunlight-driven sublimation of a once-thick mega-seasonal $CH_4$ ice deposit, producing features analogous to penitentes, albeit far larger than the water ice equivalent on Earth (A. Parker, pers. comm. 2015; Moore et al. 2017a; Moores et al. 2017).

Two large structures on the encounter hemisphere may be cryovolcanic in origin (Moore et al. 2016). Wright and Piccard Montes are enormous, roughly conical mounds featuring unusual hummocky textures. They tower 4-6 km above the surrounding landscape and extend up to ~200 km across, comparable in volume to the largest Hawai'ian shield volcanoes. Both have summit depressions at least 5 km deep. Few impact craters are superposed on them, suggesting they have been active relatively recently in Pluto's geological history.

To avoid rapid viscous relaxation, these tall structures must be made of a sturdy material like $H_2O$ ice, but there is frustratingly little information on their bulk composition, since the area appears to be mantled with an optically thick veneer of $CH_4$ ice, likely a seasonal deposit. Various cryovolcanic scenarios may be responsible for their construction (Moore et al. 2016). These include rising plumes of hot material delivering heat from the deep ice mantle, potentially involving $H_2O$ in combination with anti-freezes like ammonia or methanol. Volatiles dissolved in ascending aqueous liquids could exsolve explosively as well (Neveu et al. 2015). More exotic mechanisms involve non-$H_2O$ subsurface volatiles such as liquid $N_2$ or $CH_4$, which could explosively transition to gas upon rising through the subsurface, possibly mobilizing the $H_2O$ "bedrock" and constructing a durable edifice from it. Analogous structures appear nowhere else on the encounter hemisphere of Pluto that was imaged at high resolution, nor on icy satellites (except perhaps Doom Mons on Titan; Moore & Pappalardo 2011), and hummocky, blocky surface texture of Wright Mons is reminiscent of funiscular ("ropy") terrain adjacent to Enceladus' active, "tiger stripe" fissures.

Pluto's volatile ices migrate across its surface on various timescales, complicating interpretation of the geomorphology. Pluto's year is 248 Earth years. Its obliquity, the angle between its equatorial plane and the plane of its heliocentric orbit is currently 119°, resulting in strong seasonal variations in solar illumination versus latitude. The mean orbital eccentricity of 0.24 also strongly influences seasons, with only about 40% as much energy available from sunlight as aphelion as at perihelion. Currently, Pluto's perihelion coincides with its equinox, but the argument of perihelion circulates with a ~2.8 million year period, such that ~0.8 million years ago, perihelion coincided with northern summer, and ~2.4 million years ago, it coincided with southern summer (Dobrovolskis et al. 1997; Earle & Binzel 2015). Pluto's obliquity oscillates between ~103° and ~128° on comparable timescales, creating Milankovitch-type insolation cycles. Volatile transport models (Hansen & Paige 1996; Spencer et al. 1997; Young 2013) and global circulation models (Forget et al. 2017) show that optically thick deposits of volatile ices can migrate on seasonal timescales, and that more substantial deposits can accumulate over the longer, mega-seasonal timescales, possibly accounting for the thick mantling and retreating scarps seen in various regions. Stern et al. (2017c) explored how these mega-seasonal cycles could elevate atmospheric pressures and surface temperatures, potentially reaching the $N_2$ triple point and allowing it to flow across Pluto's surface as a liquid during mega-seasonal extremes.

Insights into seasonal migration of Pluto's volatile ices come from large-scale compositional patterns (Grundy et al. 2016a; Protopapa et al. 2017; Schmitt et al. 2017). At equatorial latitudes on the encounter hemisphere, little volatile ice is seen, apart from Tombaugh Regio and localized high altitude $CH_4$ deposits. These latitudes may remain mostly free of volatile ices year-round, since they never experience the long, dark polar winters that enable substantial seasonal deposition (Binzel et al. 2017). A belt of dark, reddish maculae girdle Pluto's equator, typified on the encounter hemisphere by Cthulhu Macula, the largest of these features seen by NH. These areas appear to be mostly devoid of exposed volatile ices and owe their coloration to organic tholins (Khare et al. 1984). Much of Pluto's tholins likely originated as photochemical haze particles, although formation within the surface ices is also possible (e.g. Materese et al. 2014, 2015). During the current epoch, haze is produced in the atmosphere at a rate that could supply tens of meters of material to the surface if continued over the age of the solar system (Cheng et al. 2017; Gao et al. 2017; Wong et al. 2017; Grundy et al. 2017). Inert $H_2O$ ice appears toward the margins of Cthulhu, and also in isolated outcrops at mid-latitudes where it is often associated with reddish material in crater floors, scarps, and mountainous regions (Cook et al. 2017b).

Pluto's northern pole has been illuminated by continuous summer sunlight for the past few decades. It is rich in $CH_4$ ice. $N_2$ and CO, Pluto's most volatile ices, are mostly seen in Tombaugh Regio, particularly in Sputnik Planitia, as well as at intermediate northern latitudes, chiefly at low-elevations where higher pressures stabilize these ices from sublimation (Grundy et al. 2016a; Protopapa et al. 2017; Howard et al. 2017b). These broad latitudinal patterns are consistent with a distillation sequence where an initially mixed volatile ice deposit progressively loses its more volatile $N_2$ and then CO components, long before the $CH_4$ is lost (Schmitt et al. 2017). This scenario is consistent with $N_2$ and CO having been mostly lost from the northern pole by the time of the encounter, but not yet from northern mid-latitudes, and they may even be still accumulating there. They are likely also accumulating at high southern latitudes currently experiencing polar winter night though this hypothesis cannot presently be observationally verified.

**2.2. Atmosphere**

NH studied Pluto's $N_2$-dominated neutral atmosphere with radio occultations, solar and stellar occultations, airglow observations, and imaging (Gladstone et al. 2016; Steffl et al. 2016; Hinson et al. 2017a; Cheng et al. 2017; Young et al. 2017). It also studied the plasma environment and the planet's solar wind interaction in situ (Bagenal et al. 2016; McComas et al. 2016). Contemporaneous observations from Earth included a ground-based stellar occultation (Sicardy et al. 2016) and ALMA observations of gaseous CO and HCN (Lellouch et al. 2016).
Post-encounter Pluto atmosphere models have to date included the variation of Pluto's atmosphere over its obliquity cycles (Stern et al. 2017c), the general circulation (Forget et al. 2017), energy balance (Strobel & Zhu 2017), chemistry (Wong et al. 2017), haze formation, sedimentation, and distribution (Gao et al. 2017, Bertrand & Forget 2017), and atmospheric escape (Hoey et al. 2017; Young et al. 2017).
From the phase delay of an uplink radio signal during the Earth occultation, Hinson et al. (2017a) derived the atmospheric pressure, temperature, and number density from the surface to ~110 km altitude (Figure 4). The ingress occultation probed the dusk atmosphere over southern Sputnik Planitia, while egress probed dawn near the sub-Charon point. Both showed strong temperature inversions, with a 3.5 km deep cold (38.9±2.1 K) boundary layer in ingress and a warmer surface temperature (57.0±3.7 K at 1 km altitude) at egress. The measured pressure was 11.5±0.7 μbar at a radius of 1189.9±0.2 km. The $N_2$ frost point at this pressure (37.2±0.1 K) is consistent with the ingress temperature, supporting the concept of a global $N_2$ atmosphere in vapor pressure equilibrium (Spencer et al. 1997). Such an atmosphere is very sensitive to the $N_2$ ice temperature, which is affected by seasons (Bertrand &

Forget 2016) and mega-seasons (Stern et al. 2017c). The analysis of the radio occultation assumed an upper boundary condition at 1302.4 km radius (~112 km altitude) that was provided by a ground-based stellar occultation (Sicardy et al. 2016), which measured temperatures from ~1191 to ~1600 km radius.

The solar occultation probed similar locations and local times as did the radio occultation. From UV spectra of transmission vs. altitude, Young et al. (2017) computed abundances of $N_2$, $CH_4$, three simple hydrocarbons ($C_2H_2$, $C_2H_4$, $C_2H_6$), and haze (Figure 5). The $N_2$ density, pressure, and temperature profiles were derived by interpolation between altitudes probed by the radio and UV occultations (~0-100 km and ~900-1000 km, respectively). The $CH_4$ mixing ratio shows a transition from eddy mixing near the surface to diffusive separation at altitude, modulated by an upward $CH_4$ flux. Young et al. (2017) constrained the homopause (i.e., where eddy diffusion equals molecular diffusion) to be at most 12 km above the surface and possibly at the surface, which is equivalent to a vertical eddy diffusion coefficient range of 550 to 4000 $cm^2$ $s^{-1}$. This slow eddy mixing is a consequence of the steep vertical temperature gradient below ~15 km altitude. The surface $CH_4$ mixing ratio was found to be ~0.28-0.35% (Young et al. 2017), maintained by warm $CH_4$-rich ices (Forget et al. 2017). Near-IR and UV $CH_4$ heating is balanced by thermal conduction in the lower atmosphere, and $H_2O$, CO, and $C_2H_2$ cooling in the middle and upper atmosphere (Zhu et al. 2014; Strobel and Zhu 2017). $CH_4$ has a long chemical lifetime, and its profile is little affected by photochemistry. However, $CH_4$ is a parent molecule for many other photochemical products (Wong et al. 2017), including $C_2H_2$, $C_2H_4$, and $C_2H_6$, all detected in the UV, and nitriles such as HCN detected in ALMA data (Lellouch et al. 2016).

Haze absorption is seen in the Alice solar occultation below ~350 km altitude (Young et al. 2017), and scattering of visible sunlight is observed in images to >200 km altitude (Figure 6; Cheng et al. 2017), with at most a few discrete cloud features (Stern et al. 2017a), all of which are low lying. The haze is forward scattering and blue in color, consistent with 0.5 μm spherical particles near the surface, and 0.4-1 μm fractal aggregates assembled from 10-20 nm monomers at ~45 km altitude (Cheng et al. 2017; Gao et al. 2017). Pluto's haze may form by a mechanism similar to Titan's detached haze, i.e., via nitrile chemistry in the ionosphere near ~500-700 km altitude where nucleation occurs (Cheng et al. 2017; Gao et al. 2017; Wong et al. 2017), followed by sedimentation, condensational growth, formation of aggregates, and further condensation of hydrocarbons that "coat" the resulting aggregates in the 200-400 km height range, and perhaps again between 5-15 km altitude. Well-ordered, globally extensive layering in the haze has been attributed to orographic gravity waves (Cheng et al. 2017).

The combined analysis of UV and radio occultations suggests a cold upper atmosphere, ~65-68 K above ~900 km (Young et al. 2017). Lighter $CH_4$ exhibits an increasing mixing ratio with altitude (Figure 5), becoming the dominant species at Pluto's ~2900 km exobase (2.5 $R_P$), where the pressure is ~0.01 pbar (Figure 4).

Molecules there undergo enhanced Jeans escape (Hoey et al. 2017; Young et al. 2017) at a rate of $(3-7)\times10^{22}$ $N_2$ $s^{-1}$ and $(4-8)\times10^{25}$ $CH_4$ $s^{-1}$. The deduced escape rate is orders of magnitude lower than predicted by pre-encounter models (Gladstone et al. 2016), because the upper atmosphere was found to be much colder than expected, possibly cooled by $H_2O$ vapor or other, as-yet unidentified mechanisms (Strobel & Zhu 2017). From the in situ plasma instruments, the deduced mass loading from the escaping gas and the ion pressure in the ionosphere together produce a $CH_4^+$ obstacle to the solar wind at ~2.5 $R_P$ on the sunward side, a bow shock ~4.5 $R_P$ sunward of Pluto, and a long $CH_4^+$ tail (>100 $R_P$), down which ~1% of the escaping gas travels (Bagenal et al. 2016; McComas et al. 2016; Zirnstein et al. 2016). A few percent of the escaping gas reaches Charon (Hoey et al. 2017) and may play a role in the production of Charon's dark, red poles (Grundy et al. 2016b). An ionospheric upper limit of 1000 cm$^{-3}$ was reported by Hinson et al. (2017b) using radio occultation measurements.

### 2.3. Interior

NH flew by Pluto at a distance of 13,700 km (11.5 $R_P$) and by Charon at 29,400 km (48.6 $R_C$), at a speed of about 13.8 km $s^{-1}$ (Stern et al. 2015). As such, it could not measure second-degree or higher-order gravity terms. Both bodies' sizes were, however, precisely measured from limb profiles on whole disk images (Nimmo et al. 2017), which when combined with masses from Earth-based astrometry (Brozović et al. 2015) yielded accurate, measured densities (see Table 1). NH lacked a magnetometer, but Pluto's solar wind interaction is consistent with that of an unmagnetized body (<30 nT at the surface; McComas et al. 2016).

The densities of Pluto and Charon are intermediate between that of rock and ice, consistent with expectations for KB bodies: subequal amounts of rock, water ice, organic matter, and volatile ices (i.e., a comet-like composition; McKinnon et al. 1997, 2008). For a composition of water ice plus anhydrous rock following solar abundances, Pluto and Charon would be about 2/3 and 3/5 rock by mass, respectively (McKinnon et al. 2017). With relatively low central pressures in Pluto and Charon (about 1.3 and 0.3 GPa, respectively), these rock mass fraction estimates are insensitive to the actual hydration state of the rock, as long as both bodies are differentiated (have rock cores and ice mantles), and indicate that Charon is slightly icier than Pluto. The organic mass fraction is unconstrained, except by comparison with comets such as 67P/Churyumov-Gerasimenko (67P/C-G), but it could be substantial (Simonelli et al. 1989; McKinnon et al. 1997; cf. Davidsson et al 2016; Fulle et al. 2016).

**2.3.1. Differentiation.** Pluto's surface shows abundant $N_2$ and $CH_4$ ice, both in Sputnik Planitia (equivalent to several 100 m in depth if distributed globally) and in the northern terrae, where local deposit thicknesses may reach 3 to 4 km as revealed by large-scale pit formation (Moore et al. 2016; Howard et al. 2017b). Methane loss to space should also be counted in this global inventory: if the 1.6 kg s$^{-1}$ loss rate above is sustained over 4.5 billion years, it would be equivalent to a 25 m global layer of methane ice. Such global inventories are evidence for differentiation of Pluto's interior (Stern 1989), but the argument is stronger for $N_2$, which was discovered a few years later than methane (Owen et al. 1993). The reason is that methane is a known cometary volatile, at the percent level or less compared with $H_2O$ (Mumma & Charnley 2011), so $CH_4$ could have been mobilized during Pluto's formation by even modest accretional heating. In contrast, comets are extremely depleted in molecular nitrogen, as exemplified by in situ mass spectrometer measurements at 67P/C-G (Rubin et al. 2015). The source(s) of nitrogen were most likely ammonia-rich aqueous lavas that erupted to Pluto's surface or nitrogen released by thermal (or hydrothermal) processing of Pluto's organic fraction. The latter requires or at least implies a rock core.

Additional support for Pluto's differentiation comes geological evidence that the outer several kilometers, or more, of both bodies are water- or volatile-ice dominated, i.e., Pluto and Charon possess low-density ice crusts.

The strongest evidence for differentiation, however, comes from tectonics. An undifferentiated interior means ice and primordial rock plus organics are mixed to the center. The water-ice melting curve cannot be intersected, as melting would separate rock from ice. This in turn requires efficient heat transport from the deep interiors of both bodies via solid-state convection, which is not implausible for water ice rheology[1]. However, radiogenic heat output was 4-to-6 times greater 4 Gyr ago than today (depending on the $^{40}K$ abundance). Over time, such declining heat flows imply cooling interiors, which for Pluto means an increasing volume of dense ice II forming at the expense of low-pressure ice I and ice V (see Figure 3b in McKinnon et al. 1997), and ice I converting to ice II within Charon. The overall effect would be a global increase in density and a gradual decrease in surface radius (for Pluto >10 km over the last 2 billion years alone) leading to pronounced compressional tectonics (e.g., thrust faults, lobate scarps)—features not seen by NH. With tectonic histories instead dominated by extension (§2.1), Pluto and Charon cannot today be undifferentiated worlds.

**2.3.2. Subsurface Ocean.** Fully differentiated structures imply ice mantles of ~300 km and 175 km thickness for Pluto and Charon, respectively (McKinnon et al. 2017; Bierson et al. 2017). Figure 7 shows conductive thermal evolution models for both bodies, accounting only for radiogenic heating. Despite minimizing melting and maximizing surface porosity by assuming a cold (150 K) start, these models readily form internal oceans. The models do not include other early heat sources such as (1) a Charon-forming impact (§ 4), (2) tidal dissipation during subsequent Pluto spin-down and Charon orbit circularization, (3) serpentinization, or (4) the potential energy released by differentiation itself. Differentiation implies ice at or near its melting point, so that after formation in a giant impact, Pluto and Charon's internal temperatures could have been much warmer than in Figure 7. The models also omit salts, ammonia, and methanol that lowers the oceans' melting/freezing points, which if included would increase their thickness and longevity. If Pluto's ocean contains ammonia, such as seen on the surfaces of Charon, Nix, and Hydra (§ 3), it would completely freeze only below the eutectic temperature of 176 K. But even with ammonia it is very hard for Charon's ocean to persist to the present day (Figure 7).

---

1 Water ice clathrate would be stable throughout both bodies, but is too rheologically stiff to allow solid-state convection (Durham et al. 2010).

Further, indirect evidence for Pluto's ocean comes from Sputnik Planitia's location, just 20° north of Pluto's anti-Charon point. Its near-alignment with the tidal axis suggests Sputnik is a positive gravity feature that drove reorientation of Pluto's figure with respect to its spin and tidal axis, i.e., True Polar Wander (TPW) to a minimum energy orientation (Nimmo et al. 2016; Keane et al. 2016). But Sputnik occupies a deep basin, forming a level plain 2.5 km below Pluto's mean elevation (Schenk et al. 2017), which suggests a mass deficit. Nitrogen ice is somewhat denser than water ice (see Scott 1976), so $N_2$ ice within the basin could contribute to a positive mass anomaly, but Nimmo et al. (2016) argue it is unlikely to be sufficient to counter the basin's overall depth (its missing mass). They propose that the basin-forming impact also resulted in the uplift of relatively dense ocean water beneath the water ice floor of the basin, much as lunar mass concentrations (mascons) are created by uplift of denser lunar mantle rocks in basin-forming impacts there (Johnson et al 2016). Nimmo et al. note that this dome of dense ocean water is subject to refreezing over time as well as infilling by warmer, less viscous $H_2O$ ice from the base of the surrounding ice shell. If, however, the ocean is cold enough (e.g., through freezing point depression), then the viscous inflow of $H_2O$ ice can be suppressed, and remarkably, the refreezing and re-equilibration time may be very long.

Globally, such cold, basal ice would imply that the lower portion of Pluto's ice shell is not convecting (Robuchon & Nimmo 2011; Hammond et al. 2016). But it cannot be too cold, because this would lead to the formation of dense ice II deep in the shell and global radial contraction, which is inconsistent with Pluto's lack of late-stage compressional tectonics (Hammond et al. 2016). Avoiding ice II formation places requirements on Pluto's subsurface temperature gradient, calling for higher heat flow from the rock core and/or lower near-surface conductivity due to volatile ices or porosity (Durham et al. 2010). Such "warm" temperature profiles also favor an internal ocean, but cannot be taken as definitive evidence for one. Larger, less dense cores, or especially, a massive carbonaceous layer at the top of the core (e.g., Figure 4b in McKinnon et al. 1997), can coexist with a warm thermal profile and no ocean.

Independent evidence that Sputnik Planitia is a positive gravity load (and thus that an ocean exists) comes from the radial orientation of multiple normal fault valleys surrounding Sputnik (Keane et al. 2016). Such a fault pattern is expected for a positive load on a spherical shell or lithosphere when the horizontal scale of the load is sufficiently large (Janes & Melosh 1990). Keane et al. (2016) attribute more distant fault and lineament patterns to a combination of stresses from large-scale TPW and global expansion due to ocean freezing. This line of argument could benefit from future attempts to establish a time history of faulting on Pluto. Keane et al. (2016) report no tectonic evidence for the despinning that would have accompanied Pluto's early, post-Charon-forming-impact tidal evolution (Barr & Collins 2015) and NH images reveal no fossil oblateness from such despinning for either body (Table 1). An ocean is consistent with if not implied by Pluto's tectonic and possible polar wander history.

## 3. Pluto's Satellites

Pluto has five known satellites: Charon, Styx, Nix, Kerberos, and Hydra, in order of their distance from the system barycenter. Charon, which is approximately half the size of Pluto, was discovered during ground-based observations in 1978 (Christy & Harrington 1978), 48 years after the discovery of Pluto. Although its mass is 12.2% of Pluto's, Charon lacks sufficient gravity to retain volatile ices (e.g., $N_2$, $CH_4$, CO) on its surface, explaining why there is no evidence for an atmosphere even at less than nanobar levels (Stern et al. 2017b). Nix and Hydra have approximately 34 times smaller diameters than Charon and were discovered during Hubble Space Telescope (HST) observations in 2005 (Weaver et al. 2006). Kerberos and Styx are smaller still, roughly one quarter the sizes of Nix and Hydra, and were discovered during HST observations in 2011 (Showalter et al. 2011) and 2012 (Showalter et al. 2012), respectively. Deep searches for additional satellites and dust rings were conducted during the NH flyby, but neither was found (Stern et al. 2015; Lauer et al. 2017).

### 3.1. Orbital Properties and Mass Constraints

The dynamical structure of the Pluto satellite system, with all five bodies in approximately coplanar circular orbits, is fascinating. Styx, Nix, Kerberos, and Hydra orbit the system barycenter, defined by the Pluto-Charon binary, with periods that are approximately integer multiples (3, 4, 5, and 6, respectively) of Charon's orbital period.

A study of the dynamical stability of orbits in the Pluto-Charon binary, before any of the small satellites were discovered (Stern et al. 1994), demonstrated the existence of stable regions that could potentially harbor satellites or rings. The discovery of Nix and Hydra motivated multiple new dynamical studies of this type (Nagy et al. 2006; Sulli & Zsigmond 2009; Pires dos Santos et al. 2011), which examined the long-term stability of the satellite orbits and pointed to additional regions of satellite stability where Styx and Kerberos were subsequently discovered. After Kerberos was discovered orbiting between Nix and Hydra, Youdin et al. (2012) showed that firm upper limits on the masses of Nix and Hydra were required to ensure the stability of Kerberos' orbit. These mass upper limits, combined with optical photometry and reasonable assumptions for the densities of the small satellites, led those researchers to conclude that Nix and Hydra must have high optical albedos, a conclusion also reached by Kenyon & Bromley (2014) and subsequently verified by the results from the NH flyby (Weaver et al. 2016; see below).

Astrometry has been used to refine the orbital elements of the satellites and place constraints on their masses (Buie et al. 2006; Tholen et al. 2008; Buie et al. 2012, 2013; Brozovic et al. 2015). The orbits of Charon, Styx, Nix, and Kerberos are circular to within the current measurement errors, but Hydra's orbit has a small, non-zero eccentricity (0.005±0.001). Additionally, the orbits of Kerberos and Hydra appear to have non-zero inclinations (0.4° and 0.3°, respectively), and none of the small satellites are in exact mean motion resonance with Charon.

Mass determinations for the small satellites are consistent with zero within the quoted errors, which rules out density determinations. Additional analysis of the astrometry and photometry of the small satellites (Showalter & Hamilton 2015) showed that Styx, Nix, and Hydra are likely tied together in a three-body resonance, reminiscent of the Laplace resonance linking Jupiter's moons Io, Europa, and Ganymede.

### 3.2. Sizes, Shapes, and Rotational Properties

Charon is a spherical body in synchronous rotation with Pluto (e.g., Stern 1992; Cheng et al. 2014a). Both images and measurements of the brightness variations of the smaller satellites show that all four are highly elongated objects rotating much faster than their synchronous rotation rates, with their rotational poles highly inclined relative to those of Pluto and Charon (Weaver et al. 2016; Porter et al. 2017). The sizes, shapes, reflectivities, and rotational properties of Pluto's satellites are listed in Table 2.

The rapid, non-synchronous rotation of the small satellites is at least partly due to Pluto's small mass, which results in >$10^9$ yr tidal timescales to achieve synchronous spin at Hydra's orbital distance (Quillen et al. 2017). Craters on the surfaces of Nix and Hydra (see later discussion) show that the small satellites suffered multiple collisions with other objects (either debris left over from the Pluto-Charon forming impact event, or from KBOs passing through the Pluto system), which may have contributed to their fast rotation periods and unusual pole orientations. However, if the small satellites were captured into mean motion resonances with Charon shortly after a Pluto-Charon creating impact event, an outward moving Charon could induce large rotational obliquity variations in the small satellites (Quillen et al. 2017).

The elongated shapes of the small satellites are typical of what is observed among the other small objects in the Solar System and presumably reflect the agglomeration of small objects into loosely bound, porous bodies whose gravity was insufficient to pull them into more spherical shapes. Kerberos has a distinctly double-lobed shape suggesting the merger of two smaller bodies, as was proposed for the similarly shaped nucleus of comet 67P/Churyumov-Gerasimenko (Massironi et al. 2015).

### 3.3. Surface Morphologies

Charon's surface displays impressive tectonic structures associated with its evolution during the first hundreds of millions of years following its formation (Stern et al. 2015; Moore et al. 2016; Beyer et al. 2017). Two huge chasms encircle much of Charon's encounter hemisphere. Serenity Chasma is more than 50 km wide, ~5 km deep, and ~200 km long. Mandjet Chasma is at least 450 km long, ~30 km wide, and reaches ~7 km deep. These chasmata were possibly produced when a primordial ocean under a surface ice shell froze and expanded, splitting the surface apart (Moore et al. 2016). Most of the northern hemisphere is extremely rugged with a number of polygonal troughs ~3-6 km deep scattered throughout. All of these morphological features resemble extensional rifts, similar to features observed on mid-sized icy satellites (Collins et al. 2010).

The regions slightly north of the equator and at southern latitudes on Charon's encounter hemisphere generally have a much smoother surface. The vast (~540,000 km$^2$) smooth plain informally named Vulcan Planum is criss-crossed with aligned rilles, which range from less than a kilometer to ~3 km in width and are ~0.5 km in depth (Beyer et al. 2017). These rilles have an arcuate or sinuous pattern and can be up to hundreds of kilometers long. The rounded lobate features in Vulcan Planum (e.g., the "moated mountains of Clarke, Kubrick, and Butler Montes) appear to be the result of flooding or other flow emplacement of icy material on the surface. Both the rilles and the lobate features are probably associated with cryovolcanism (Beyer et al. 2017). The extent of the resurfacing on Charon is comparable to that also apparently seen on several Saturnian and Uranian satellites (e.g., Ariel, Dione, and Miranda). The cryovolcanic activity on Charon might be internally driven, presumably by a combination of interior heating associated with the collisional formation event and natural radioactivity; tidally driven heating during the early orbital evolution of Pluto and Charon is another possibility.

Both Nix and Hydra appear to have relatively smooth surfaces at spatial scales of ~0.3 km and ~1 km for Nix and Hydra, respectively (Figure 8). This interpretation may, however, be biased by the low phase angles of most of the images obtained. For Nix at least, higher phase angle imaging hints at a more rugged surface (see figure S13 in Weaver et al. 2016).

Crater statistics show that Charon's surface is ancient (~4 Gyr) but with some signs of more recent activity (Moore et al. 2016). Crater statistics for Nix and Hydra indicate ancient surfaces for those bodies as well (Weaver et al. 2016), dating binary and satellite system formation to be early in the history of the solar system (Stern et al. 2015). In Charon's case, where crater counts on Vulcan Planum are complete down to ~1-km diameter, the number of small craters is well below the number expected prior to the NH encounter. This provides strong evidence that the KB size frequency distribution (SFD) rolls over at small sizes much more rapidly than predicted by current models (Singer et al. 2017; Greenstreet et al. 2015). apparently either some unknown process depleted the small end of the KBO size distribution, or the observed shallow SFD at small sizes is a relic of the accretionary epoch (Singer et al. 2017).

### 3.4. Albedos

Charon's albedo variation is much less extreme than Pluto's, with normal reflectances of 0.20 to 0.73 (Buratti et al. 2017). Charon's globally averaged albedo is 0.41±0.02 (derived from data presented in Buie et al. 2010a), which is significantly lower than Pluto's (0.52±0.03; Buratti et al. 2015). This is explained by the lack of extensive mobile (hence fresh) surface volatiles ($N_2$, $CH_4$, $CO$), which are howecver present on Pluto (Stern et al. 1987).

Surprisingly, the visible wavelength albedos of Pluto's small satellites are significantly higher than Charon's, and comparable to, or higher even than Pluto's (Weaver et al. 2016). The vast majority of small KBOs (but still usually larger than 100 km in diameter, we note) for which the relevant observational data are available have inferred V-band albedos <20%, with typical values of ~10% (Vilenius et al. 2012; Lellouch et al. 2013). The high albedos of Pluto's small satellites, which cannot be due to condensed surface volatiles (they are far too small to retain those), is a puzzle.

### 3.5. Surface Colors and Compositions

Charon's surface geomorphological, compositional, and reflectivity variegation are not nearly as diverse as Pluto's (Moore et al. 2016; Grundy et al. 2016a). Charon's surface is generally rather gray, consistent with its surface being dominated by water ice, but there is a northern polar hood (informally named Mordor Macula) that is darker and redder than the rest of the surface (Figures 9, 11). This polar hood can be explained by a remarkable process not seen anywhere else in the Solar System (Grundy et al. 2016b). Briefly, methane ($CH_4$) molecules, the dominant species escaping Pluto's atmosphere, are preferentially captured onto the coldest regions of Charon's surface (the poles) where irradiation (e.g., from ultraviolet light and cosmic rays) processes it into more complex, thermally involatile hydrocarbons that are darker and redder than the original condensed $CH_4$.

The surfaces of Nix and Hydra are also rather gray, as they too have compositions dominated by water ice (see below). Their surface colors appear to be less diverse than Charon's, but that may be a selection effect because there were so few pixels in the NH images of Nix and Hydra. Hydra's surface color is slightly bluer than Nix's, which is consistent with Hydra's higher visible light albedo. The color camera on NH did not resolve Styx and Kerberos, owing to their large distances from the spacecraft, but their global colors are gray.

Remote observations from Earth at near-infrared wavelengths have long demonstrated that water ($H_2O$) ice is the dominant constituent of Charon's surface (e.g., Buie et al. 1987). Global near-infrared spectra have also consistently identified an absorption near 2.2 µm, which has usually been attributed to an ammonia-bearing species (e.g., ammonia hydrate, ammonium hydrate, or mixtures of $NH_3$ and $H_2O$ ices; e.g. Cook et al. 2007). NH observations show that the 2.2 µm feature can be marginally detected over much of Charon's surface, but there are specific areas where the absorption is enhanced, for example near the relatively bright Organa crater (Figure 9). If the 2.2 µm feature is indeed attributable to $NH_3$-bearing species, these regions of enhanced abundance may indicate relatively recent exposure of formerly buried material because the time scale for radiolytic destruction of ammonia ice on Charon's surface is only ~$10^7$ years (Grundy et al. 2016a).

Near-infrared spectral data taken during the NH flyby confirmed that water ice is the dominant surface component of Nix, Hydra, and Kerberos (Cook et al. 2017a). The Nix and Hydra spectra also show an absorption near 2.2 µm, similar to the feature seen near the Organa crater on Charon (Figure 9). Since the crater retention ages of Nix and Hydra are ancient, and since cryovolcanism is very unlikely to operate on such small bodies, the presence of ammonia-bearing ices on their surfaces is puzzling, like their high albedos.

## 4. Origins

### 4.1. Pluto's Origin

Pluto's eccentric and inclined, resonant orbit, and the later discovery of other, much smaller KBOs in the 3:2 and other mean-motion resonances with Neptune, were the first, major clues that Neptune's orbit had migrated outward (Malhotra 1993, 1995). Difficulties in explaining Pluto's 17° inclination, and the wide range of Plutino inclinations generally (among other dynamical indicators), ultimately led to consideration of migration of all 4 giant planets due to energy and angular momentum exchange with an exterior planetesimal disk of some tens of Earth masses: the now well-known "Nice model" of giant planet migration/instability (Tsganis et al. 2005; Levison et al. 2008) and subsequent variations and elaborations.

The instability in the orbits of the original giant planets, which allows for a profound rearrangement of global solar system architecture (and formation of the KB, among other things), is a defining aspect of the Nice model. Initial conditions, if carefully chosen, can also delay the instability by hundreds of millions of years, and so in addition potentially provide an explanation for the Late Heavy Bombardment of the Moon and terrestrial planets (Gomes et al. 2005; Levison et al. 2011). Such a late instability is not, however, an intrinsic dynamical feature of the Nice model. Most Nice-like numerical simulations go unstable on a shorter time scale and involve ejection of a giant planet (specifically one of the ice giants Uranus or Neptune). Accordingly, more recent orbital migration/instability models of this type have posited 5 (or 6) original giant planets, one (or 2) of which are lost in the first tens of millions of years after dispersal of the solar nebula (e.g., Nesvorný & Morbidelli 2012). These later models have proven remarkably successful in explaining a wide variety of dynamical features of the Solar System, particularly the orbital structure of the KB: its various resonant populations (of which the 3:2 is only one of many), the hot *and* cold classicals, the broader, scattered/scattering disk (bodies in distant orbits still interacting gravitationally with Neptune), and the detached population (those non-resonant bodies that do not interact with Neptune) (Nesvorný 2015; Nesvorný & Vokrouhlický 2016; Nesvorný et al. 2016).

All of the KB populations above, save for the cold classicals, are believed to have originated in the remnant planetesimal disk whose outer edge was near 30 AU (otherwise Neptune would have migrated farther via its dynamical interactions with the disk). Pluto accreted in this disk and was likely one of one to several thousand similar or greater mass bodies (Stern 1991; Nesvorný & Vokrouhlický 2016). The corresponding implantation efficiency into the KB of $10^{-3}$–$10^{-4}$ is consistent with the existence of Pluto and Eris and other large DPs in the KB (and presumably the Oort Cloud), and Neptune's capture of Triton from the same remnant planetesimal disk (e.g., Nogueira et al. 2011).

The remnant planetesimal disk, ranging from perhaps 20 AU to 30 AU from the Sun, was also the birthplace of what are today's short-period (or Jupiter-family) comets and Centaurs. As such, the icy volatiles seen on Pluto-Charon ($CH_4$, $N_2$, CO, $NH_3$) and its rock-rich nature (§2.3) are consistent with cometary chemistry and comets' dust-rich composition (Mumma & Charnley 2011; Davidsson et al. 2016). Notable differences are the much higher $N_2$/CO ratio seen on Pluto than in cometary comae (discussed in 2.3.1), and the non-detection of $CO_2$ ice, a common cometary volatile, on either Pluto or Charon. The infrared spectral signature of $CO_2$ ice at 2 μm is clear on Triton (Grundy et al. 2010), and is also seen on several Uranian satellites, whereas the apparent absence of $CO_2$ throughout the Pluto system, even at the high spatial resolution afforded by NH, is puzzling. Perhaps $CO_2$ in comets is a radiolysis product in the same manner that the molecular $O_2$ discovered in 67P/CG's coma may be (Bieler et al. 2015), noting that $O_2$ was *not* detected in Pluto's atmosphere (§2.3). The strong $CO_2$-ice fundamental at 4.28 μm, outside the LEISA spectral range, may be observed in the future on Pluto and/or Charon by JWST.

The densities of both Pluto and Charon are close to but less than 2000 kg m$^{-3}$ (Table 1), somewhat less than that of Triton (2060 kg m$^{-3}$), and clearly less than the values for those comparably large dwarf planets for which we have reliable densities (Eris and Haumea; 2500–2600 kg m$^{-3}$) (McKinnon et al. 2017). Haumea almost certainly lost water ice in a massive collision (e.g., Barr & Schwamb 2016), and so was originally more Pluto-like in terms of density. The overall rockiness of outer solid, heliocentric bodies has long been tied to the Solar System's C/O ratio and sequestering of oxygen by CO (thus reducing the $H_2O$-ice abundance) (McKinnon & Mueller 1988; Stern et al. 1997b; Wong et al. 2008).

As to the accretion of Pluto itself (or its progenitors, see below), it had long been thought a significant conundrum, in that accretion times in the distant outer solar system would be exceedingly long (Stern et al. 1997b). Moving Pluto's formation region to 20-30 AU largely mitigates the problem. Kenyon & Bromley (2012) show that standard hierarchical coagulation simulations can produce Pluto-class bodies in tens of millions of years, provided a minimum-mass solar nebula of solids is assumed and that the planetesimals start small, but not too small (0.1–1 km).

Forming such smallish planetesimals in the gas disk may, however, not be easy. Various dynamical barriers must be overcome, and such difficulties were a principal rationale that drove the discovery and theoretical development of the streaming instability (SI). In SI, variations in particle ("pebble") number density in the protoplanetary gas disk self-amplify through drag and pressure forces (Youdin & Goodman 2005), ultimately leading to gravitational collapse of relatively massive "pebble piles" — i.e., *big* planetesimals can form quickly (see Johansen et al. (2014) for a recent review). Simulations (Johansen et al. 2015) suggest that prompt, initial formation of planetesimals up to ~100-km radius is possible at ~25 AU in the solar nebula by SI, with growth to Pluto-scale and larger in a few million years by further pebble accretion.

## 4.2. Satellite Origins

Binaries—or more generally, satellites—are common in the KB (Noll et al. 2008; Fraser et al. 2017), and satellites are essentially ubiquitous among nearly large KBOs (>1000-km diameter). Taking Pluto-Charon as a prime example, three general mechanisms have been proposed or can be considered for binary formation: (1) a massive collision between sizeable precursors (McKinnon 1984, 1989; Canup 2005, 2011); (2) gravitational collapse and fission of a massive, rotating pebble cloud formed by SI in the solar nebula (Nesvorný et al. 2010); and (3) dissipative gravitational interactions between "large" KBOs in the presence of a massive, dynamically cold, small body (or pebble) disk (Goldreich et al. 2002; Schlichting & Sari 2008).

We discuss mechanism (1) in the next section.

Mechanism (2) requires a gravitationally unstable pebble cloud as least as massive as Pluto-Charon, because mass is typically lost during binary formation due to scattering (Nesvorný et al. 2010). Simulations shown in Johansen et al. (2015) suggest that the formation of particle clouds of this mass scale is unlikely. Furthermore, gravitational collapse in SI typically leads to formation of higher multiplicity systems (such as a binary orbited by a more distant companion, as in star formation), and it is not clear that an outer satellite naturally ends up in the orbital plane of the central binary, much less if a family of coplanar small satellites can form. Mechanism (2) does predict that all members of the system should share the same initial composition, which is *not* consistent with Charon's lower rock/ice ratio compared with Pluto (McKinnon et al. 2017) or with the apparent iciness of the small satellites, based on their high albedos and strong $H_2O$-ice spectral signatures (Weaver et al. 2016).

Mechanism (3) requires an enormous mid-plane pebble reservoir to provide the dynamical friction to damp the larger KBO encounter velocities. Goldreich et al. (2002) postulated an initially bimodal size distribution of planetesimals in the primordial disk with $\sigma/\Sigma \sim 10^3$, where $\sigma$ and $\Sigma$ are the surface densities of pebbles and 100-km class bodies, respectively, a large factor that Nesvorný et al. (2010) note is necessary for efficient binary formation. However, such a ratio is unlikely to be maintained as the larger KBO cohort grows to Charon size and larger, and $\sigma/\Sigma$ likely ultimately declines to 10 or less (Nesvorný & Vokrouhlický 2016). Likewise, if the end stages of Pluto's accretion involve gas-free, hierarchical collisions, then gravitational stirring would all but eliminate mechanism (3).

Simultaneous growth by pebble accretion onto an existing small, wide binary produced by either mechanisms (2) or (3) cannot produce a Pluto-Charon either, because the requisite angular momentum cannot be provided (Stern et al. 1997b). In addition, the highly oblique nature of the Pluto system is not predicted by mechanisms (2) or (3) (Schlichting & Sari 2008), whereas it is a natural consequence of a large collision (mechanism 1).

### 4.3. Formation of the Pluto System by Giant Impact

Here we focus on the leading mechanism for the formation of the Pluto system: giant impact (e.g., Canup 2005, 2011; Sekine et al. 2017). The formation of the Pluto system by giant impact in many ways resembles scenarios for the formation of the Earth's Moon, with the proviso that the scales, velocities, and densities are proportionally smaller, which have important consequences.

Low approach speeds ($v_\infty < 1.2\ v_{esc}$) are a general prerequisite to launching substantial mass into permanent orbit (McKinnon 1989; Canup 2005, 2011), which is consistent with the presumed dynamical conditions in the remnant planetesimal disk. This "ancestral KB" would have been at least 100 times as massive as the KB today, making collisions between such large protoplanets likely, if not inevitable (Stern 1991). In the example illustrated, "Charon" is derived largely intact from a *portion* of the impactor (Canup 2011), a physical result that differs from the so-called intact captures illustrated in Canup (2005). The icy shells are ejected, and fragments rain back down on the re-accreting Pluto and Charon (Smullen & Kratter 2017), while other ice-rich debris forms an initially dynamically hot disk that extends well beyond Charon's orbit. Some portion of this disk is then available to form the small, exterior satellites in the same plane as Charon's orbit (Kenyon & Bromley 2014; Bromley & Kenyon 2015).

If the precursors have no icy shells, and both bodies are initially undifferentiated, a Charon can still form, but it will necessarily have the same composition as Pluto in terms of ice/rock, and typically there will be no satellite-forming disk (Canup 2005). Such simulations do not match observations of the Pluto system, in that Pluto appears more rock-rich than Charon (Section 2.3) and possesses an extensive satellite system. Conversely, if the precursor bodies are initially fully differentiated (e.g., as in Figure 7), then the outcome of collisions is an ice-rich disk from which a lower-density, ice-rich Charon (and presumably smaller satellites) form (Canup 2011). This is clearly ruled out by Charon's density (which implies a fairly high rock mass fraction; see Table 1). Hence, the impact simulations between initially *partially differentiated* precursor bodies are the most successful to date in generating a Pluto-Charon-like binary while also producing apparently ice-rich (if not pure ice) small satellites (Weaver et al. 2016)—the last a feature predicted in Canup (2011).

We note that the recent giant impact simulations of Sekine et al. (2017) only considered undifferentiated impactors, and so did not generate small-satellite-forming disks. The focus of these authors, however, was on thermal effects of the Charon-forming impact. Unlike the Earth-Moon case, the thermal effects for Pluto and Charon are generally modest due to the relatively slow impact speeds involved. Volume averaged temperature changes do not exceed 100 K (McKinnon 1989; Canup 2005), but the distribution of impact heating is nonuniform, and can easily cause localized water-ice melting and vaporization of volatile ices (Canup 2011; Sekine et al. 2017). Starting from warm, already partially differentiated precursors, it is likely that the Charon-forming impact was the trigger for full differentiation of Pluto (McKinnon et al. 1997). The situation for Charon is less clear. The satellite that forms is essentially unheated in most simulations (Canup 2011; Sekine et al. 2017), but SPH models to date do not include strength or material friction. When strength and dissipation are included in future models, the resulting post-impact temperature patterns are likely to change (Davies & Stewart 2016).

**4.3.1. The Small Satellite Conundrum.** The successful numerical simulations in Canup (2011) yield debris disks external to Charon's orbital position, but these are generally much more compact than the current positions of Styx, Nix, Kerberos, and Hydra (whose semi-major axes range from ∼36 to ∼55 $R_P$). Presumably, a small fraction of this debris was captured into resonances with Charon, and each other, when Charon was close to Pluto, which would prevent collisions with Charon as the large moon and its retinue of small satellites tidally evolved outward from Pluto (e.g., Stern et al. 2006). As attractive as this formation picture is, various attempts by multiple researchers (Ward & Canup 2006; Cheng et al. 2014b; Walsh & Levison 2015) have failed to create a viable model that starts with a relatively compact grouping of six objects (including Pluto) produced soon after the impact event, and which then dynamically evolves to the system observed today. And no models to date have attempted to model the formation of the putative Laplace resonance between Styx, Nix, and Hydra. Nevertheless, the giant impact hypothesis remains the favored formation scenario, because later independent capture of multiple objects from the KB by Pluto, which is another potential formation mechanism, is highly unlikely to produce the coplanar, closely spaced satellite orbits observed today (see review by Peale & Canup 2015).

It has also been suggested (Lithwick & Wu 2008) that Nix and Hydra formed within a collisional plutocentric disk composed of small bodies captured from heliocentric orbits. The high albedos and icy surface compositions of the small satellites are, however, more consistent with bodies formed from the icy mantles stripped from two large differentiated colliding bodies, and much less with primitive, dark (low-albedo) objects captured from the general KB population, which argues against this hypothesis.

Kenyon & Bromley (2014) and Bromley & Kenyon (2015) argue that the original, impact-derived, compact debris disk could have spread viscously out to and beyond the present position of outermost Hydra, and predicted that distant small (up to several-km wide) satellites await detection there. Unfortunately this size is just at or below NH LORRI detection limits, and no new external satellites >1.7-km wide were seen (for a Nix-like albedo of 0.5; Weaver et al. 2016). Alternatively, it may be that whatever small satellites initially formed after the giant impact are not the satellites we see today. Walsh & Levison (2015) advocate that one or more destructive collisions with heliocentric impactors have remade (and expanded) the small satellite system. The apparent composite nature of Hydra and Kerberos (§3) indeed suggests that the full history of the small satellites may involve stochastic events such as resonance formation and loss, chaotic orbit destabilization, impact breakup, and reformation. Styx, Nix, Kerberos and Hydra may simply be the survivors of a complex, contingent history, a history that by its very nature will resist full revelation.

## 5. Implications for Small KB Planets

Our greatly increased knowledge of the Pluto system after its reconnaissance by NH can be translated into new perspectives on the entire class of small KB planets that Pluto represents. We provide here emergent expectations for the future groundbased and spacebased exploration of this important population:

- **Expect Surprises**. If the flyby of Pluto taught us anything, it reemphasized the lesson of many other first terrestrial planet flybys: one simply cannot predict the richness of nature and the range of physical expressions on far away worlds without the resolution and in situ presence that close up exploration enables.

- **Expect Activity**. We discovered that Pluto has been highly geologically active on a variety of scales, throughout its history, and without an ongoing tidal energy source. We should not be surprised to find similar or greater degrees of activity on small planets at large or larger heliocentric distances. Owing to higher rock fractions (Sicardy et al. 2011) and therefore greater radiogenic reservoirs, this might be particularly common in higher density bodies with icy surfaces like Eris.

- **Expect Complexity**. The enormous range of geological expression, compositional diversity, and atmospheric/climate phenomena that are expressed on Pluto may well be equally common on similarly sized bodies elsewhere in the KB and OC.

- **Expect Diversity**. It is already known that the small planets of the KB express a significant range of bulk densities, satellite system configurations, surface compositions, colors, and albedos. We believe that this finding, as well as the completely different appearances and variegated atmospheric states and surface compositions seen among the three KB worlds visited to date by spacecraft (Triton, Pluto, and Charon), are telling us that planets like Haumea, Makemake, Quaoar, and Eris are likely to constitute a highly heterogeneous population, perhaps even more so than the terrestrial planets.

- **Expect Sputnik Planitia Analogs**? Pluto's high albedo and compositionally $N_2$-dominated, thermally convecting surface unit called Sputnik Planitia may have analogs on other KB and OC planets. Indeed, Eris is known to exhibit a globally high albedo and volatile ice surface composition (Licandro et al. 2006a,b; Dumas et al. 2007; Tegler et al. 2010) and we speculate that it could in the extreme be a globally convecting, "Sputnik Planitia world."

- **Expect $NH_3$**. The surprising finding of $NH_3$ and/or $NH_3$-bearing compounds on all of the satellites in the Pluto system that NH studied compositionally shows that ammoniated species are more stable on surfaces in the KB than had been expected. That in turn suggests $NH_3$ may occur elsewhere on bodies across the KB and, possibly, in the OC. As demonstrated by its presence on tiny Nix, $NH_3$'s presence clearly does not require cryovolcanism from deep interior reservoirs.

- **Expect Climate Change and Consequent Atmospheric and Surface Temporal Variability.** We summarized evidence for Pluto's strong seasonal, orbital, and mega-seasonal climate (i.e., atmospheric temperature and pressure) variations owing to the combined effects of its high obliquity, its orbital and axial precessions, and its elliptical orbit and the exponential dependence of the vapor pressures of its surface $N_2$, CO, and $CH_4$ ices on temperature. Some other small planets in the KB are already known to display $N_2$ and $CH_4$ on their surfaces and a wide variety of orbital eccentricities and obliquity states indicating to us that similar or even greater degrees of climate variation and volatile migration may be expected there. As such, the apparent rarity of KB planet atmospheres that we observe today could be simply a temporal coincidence as the individual bodies vary in their long-term atmospheric cycles.

- **Expect Evidence of Liquids.** We have referred to evidence for past episodes of standing or flowing liquids/slurries on Pluto's surface. Given this and that numerous other KB planets display $N_2$ and $CH_4$ on their surfaces, combined with the qualitative arguments presented just above to expect strong climate cycles on many of those worlds, suggests to us that other KB planets may also reveal evidence for previous epochs of standing or flowing liquids or slurries on their surfaces.

- **Expect Interior Oceans.** Above we reviewed the evidence that Pluto may well harbor an interior ocean and that Charon might have had the same in its youth. Given the ubiquity of $H_2O$-ice in the outer solar system and the similar pressure and temperature regimes at depth in other DPs, we suspect that many of the larger members of Pluto's cohort in the KB and OC harbor interior oceans as well.

## 6. Pluto Follow On Missions

Despite the great advances made possible by NH, many aspects of the Pluto system remain difficult to understand. We do not foresee many important new advances unless a return mission is sent to orbit Pluto and study the system in more detail.

The exploration of Pluto and its satellites by a future orbiter will reveal the geology and composition of the remaining ~60% of Pluto and Charon that were not mapped in detail. Such an orbiter will also allow the first detailed, close range studies of Pluto's small satellites, and the study of how Pluto's atmosphere and surface changes with time. Such a mission, if properly equipped with a comprehensive payload, would address numerous open scientific issues. For example, it would allow the nature of Pluto's likely interior ocean and its interior structure to be determined, the formation mechanism of Pluto's many haze layers to be explained, and a comprehensive search for volcanic and geyser-like activity to be undertaken.

Such an orbiter should carry high resolution panchromatic and color imagers, an atmospheric mapping UV spectrograph, and an infrared surface composition mapping spectrograph, and radio science instrumentation, as NH did. Such a mission would also benefit from carrying an atmospheric mass spectrometer, a magnetometer, a ground penetrating radar, a surface mapping LIDAR or laser altimeter, a nephelometer, a sensitive Doppler gravimetric mapping capability, and thermal mapping capability.

Also of high interest for future KB exploration are reconnaissance flybys with similar capability as NH to sample a diversity suite of other KB planets. Recently, we have discovered that a choice between a Pluto orbiter and additional KB planet reconnaissance missions need not be made. In fact, it now appears feasible that a combined Pluto orbiter/KB reconnaissance flyby mission can be implemented using current technology electric propulsion fed by Radioisotope Thermoelectric Generator power sources like those that enabled NH. Such a mission, in our estimate, could powerfully compete with other proposed 2020s and 2030s planetary exploration missions for scientific merit and feasibility and should be considered by the next Planetary Decadal Survey.

## 7. Acknowledgements

We thank all the members of NASA's NH team, the members of the 2003 Planetary Decadal Survey, the members of the 1996 Pluto KB Mission Science Definition Team, and NASA for making possible the exploration of Pluto and the initial reconnaissance of the KB by spacecraft. We also thank the many groundbased observers who have worked on the exploration of the Pluto system and the KB for their work to uncover the secrets of the ninth planet and the KB from afar. We thank Dale Cruikshank, Carey, Jonathan Lunine, Jeffery Moore, Francis Nimmo, Mike Summers, and Darrel Strobel for their very helpful reviews of this paper. This work was funded by the NASA NH project.

We dedicate this paper to Clyde W. Tombaugh, who discovered the third zone of the solar system two generations ahead of his time.

**Figure Captions**

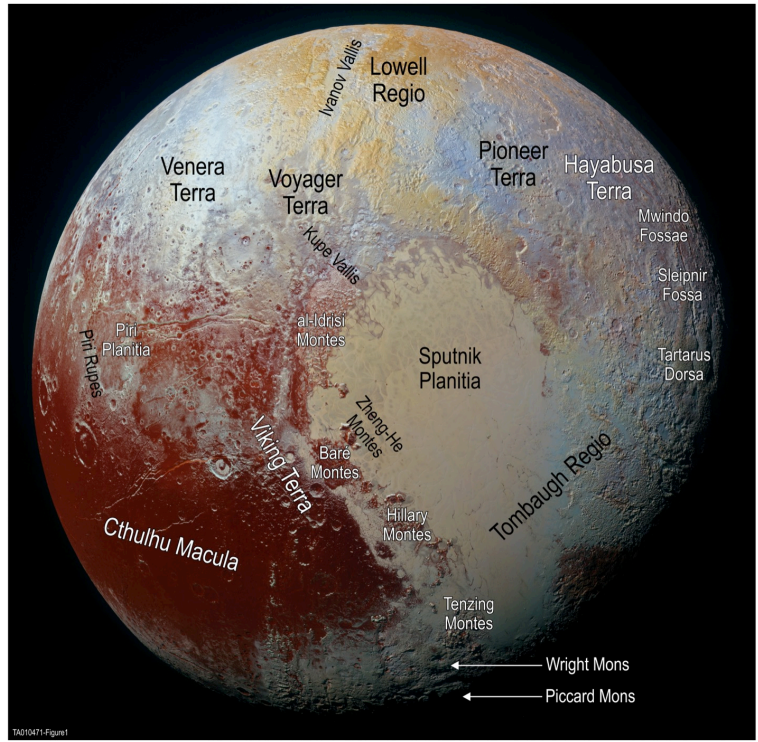

**Figure 1.** Pluto's encounter hemisphere as seen by NH in enhanced color, produced by two visible wavelength filters (400-550 and 540-700 nm) and one near-infrared filter (780-975 nm) shown in blue, green, and red colors, respectively. Formal and informal place names mentioned in the text are indicated.

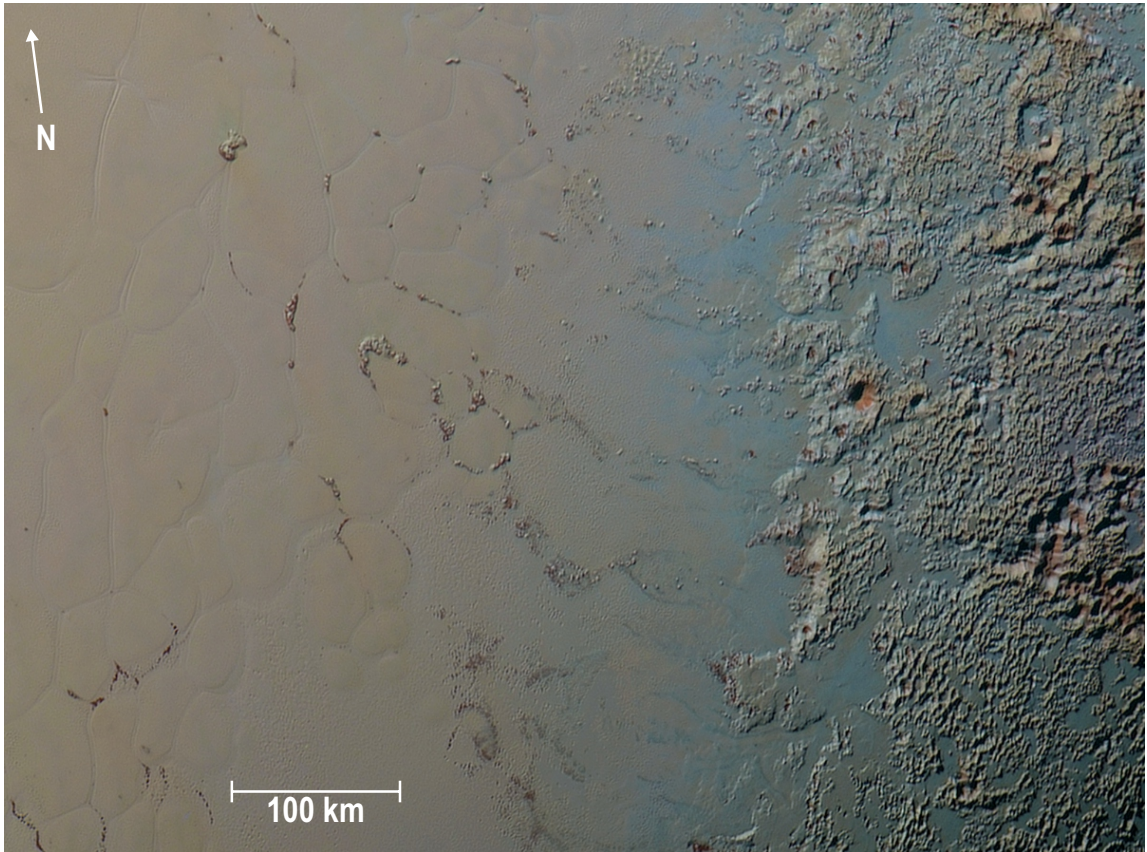

**Figure 2.** Transition between cellular terrain in glacial ices in eastern Sputnik Planitia (at left) and pitted uplands in eastern Tombaugh Regio (at right) with valley glaciers draining west into Sputnik. Chains of small hills, likely constructed of $H_2O$- and/or $CH_4$-ice trace glacial flow streams into Sputnik, tending to converge on downwelling margins of convection cells.

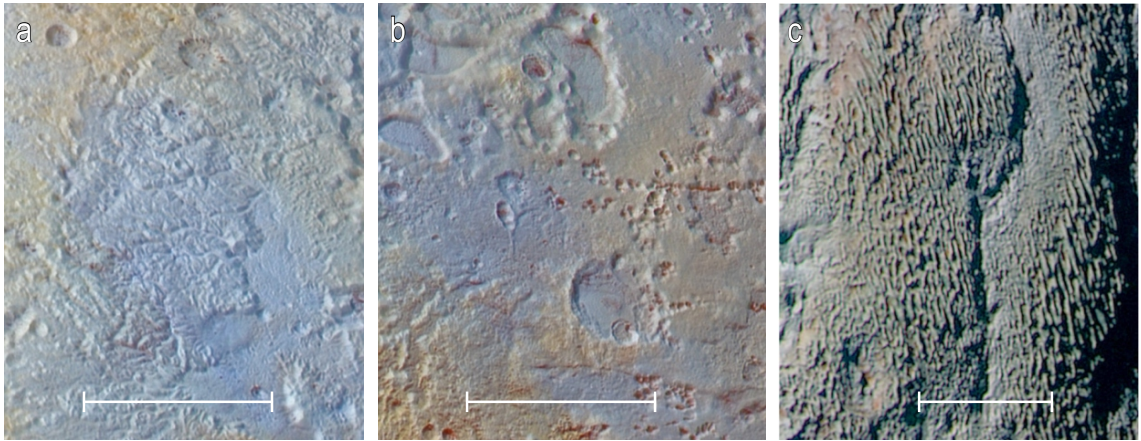

Three of Pluto's distinctive geomorphologies. (a). Dendritic valley networks in Pioneer Terra. (b). Irregularly shaped, flat-floored pits at upper left, and pit clusters at lower right, also in Pioneer Terra. (c). Bladed terrain in Tartarus Dorsae. Scale bars are each 100 km.

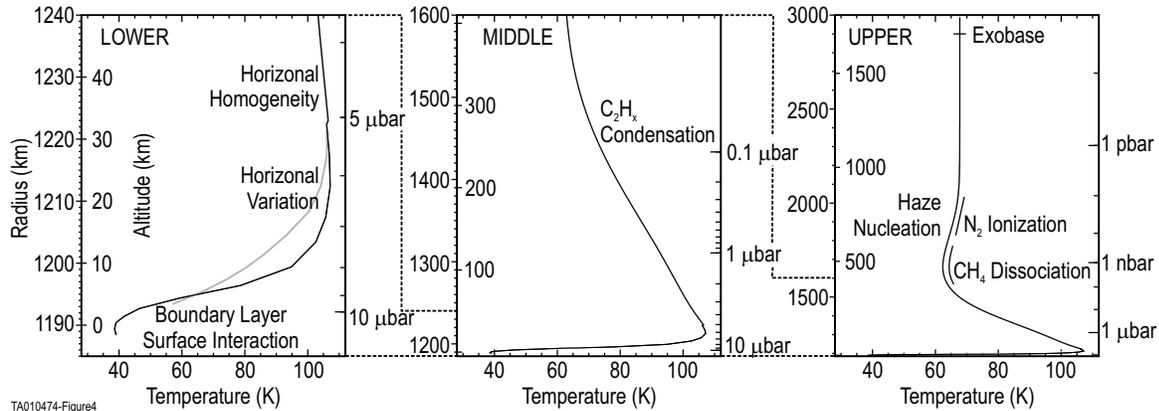

**Figure 4.** Overview of Pluto's vertical atmospheric structure at the time of the NH flyby. Below 30 km, black indicates the dusk atmosphere over the southern tip of Sputnik Planita (radio ingress), while gray indicates the dusk atmosphere over an uncertain terrain type near the center of the Charon-facing hemisphere (radio egress). Above 30 km, ingress and egress are similar in both the radio and UV occultations. The surface at radio ingress was $R_P$=1187.4±3.6 km, while at egress $R_P$=1192.4±3.6 km. The altitude scale adopts $R_P$=1190 km. Figure from Young et al. (2017).

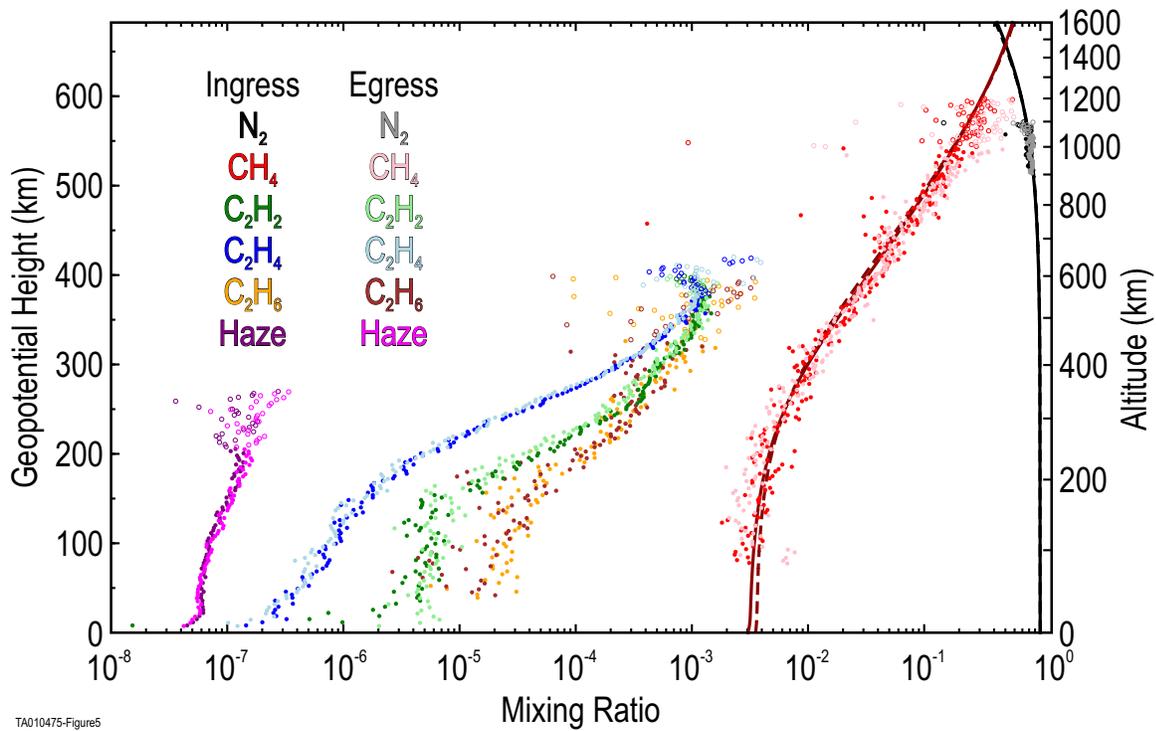

**Figure 5.** Mixing ratio versus geopotential height (left axis) and altitude (right axis) derived from the Alice solar occultation. For haze, the value plotted is $\varepsilon*10^{15}$ cm$^2/n_{total}$, where $\varepsilon$ is the extinction coefficient in cm$^{-1}$. Solid lines are two models of CH$_4$ vertical transport. Open circles indicate where the uncertainty exceeds a factor of 2.7 (e). Figure adapted from Young et al. (2017).

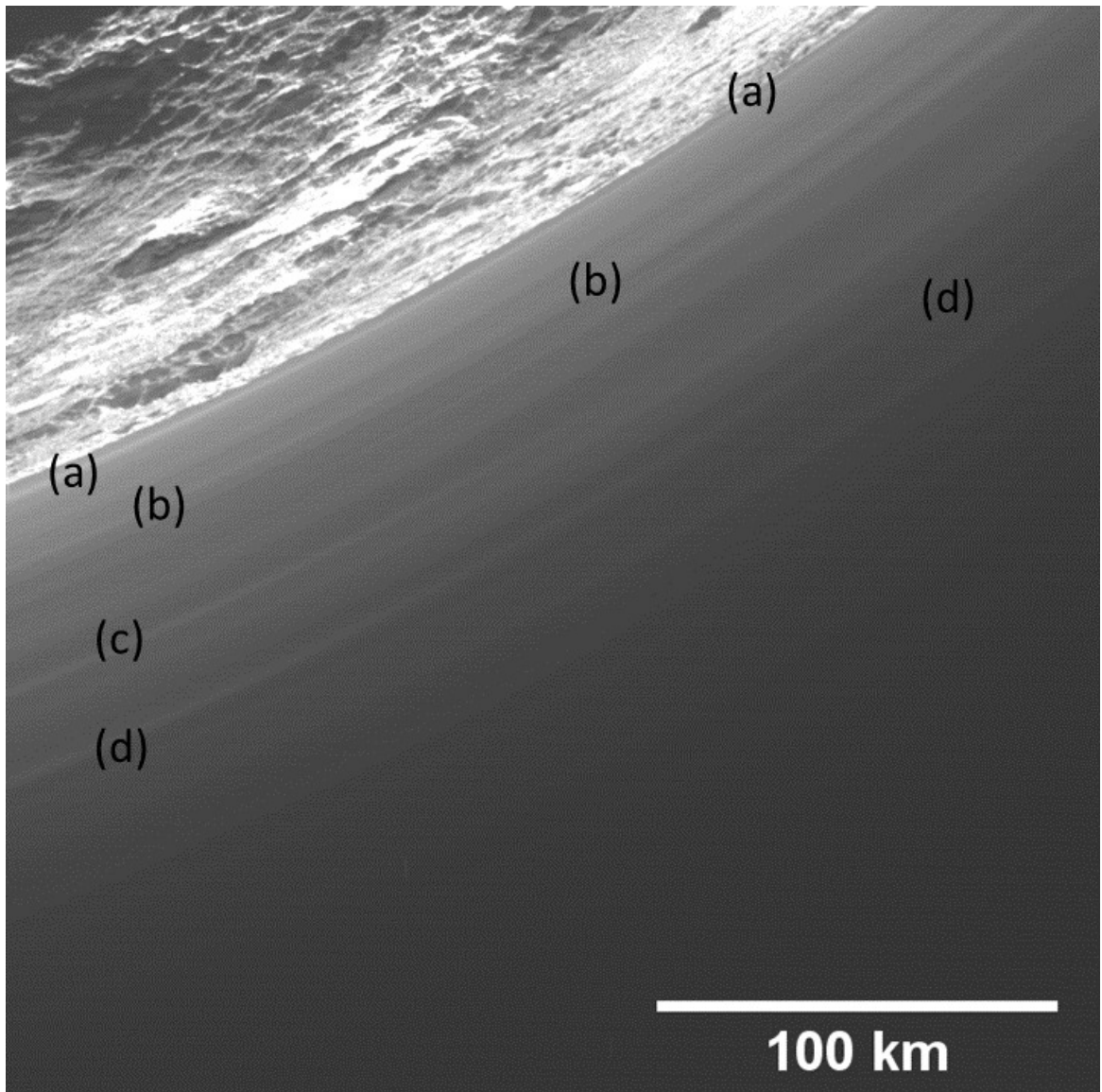

**Figure 6.** Pluto's haze is measurable to greater than 200 km above the surface in this panchromatic image at phase 147°. Haze layers horizontally change thickness (a), merge/split (b), or appear/disappear (c). A dark lane at altitude 72 km, at (d), is near the minimum of the atmospheric buoyancy frequency. Figure from Cheng et al. (2017).

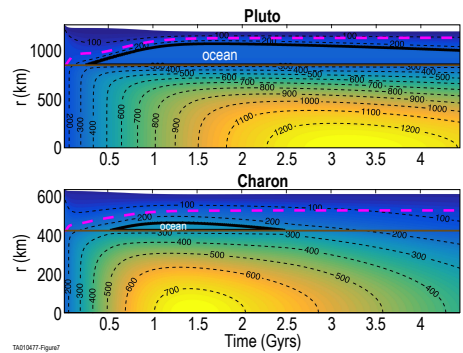

**Figure 7.** Modeled conductive thermal histories of Pluto and Charon, adapted from Bierson et al. (2017). Rock cores underlie initially porous ice mantles, and both respond conductively to radiogenic heat release ($^{238}$U, $^{235}$U, $^{232}$Th, $^{40}$K) in the core. Isotherms are labeled in K, with ice melting at depth (internal oceans) indicated. Pink dashed lines indicate the limits of porosity, which is progressively eliminated in warmer, softer ice.

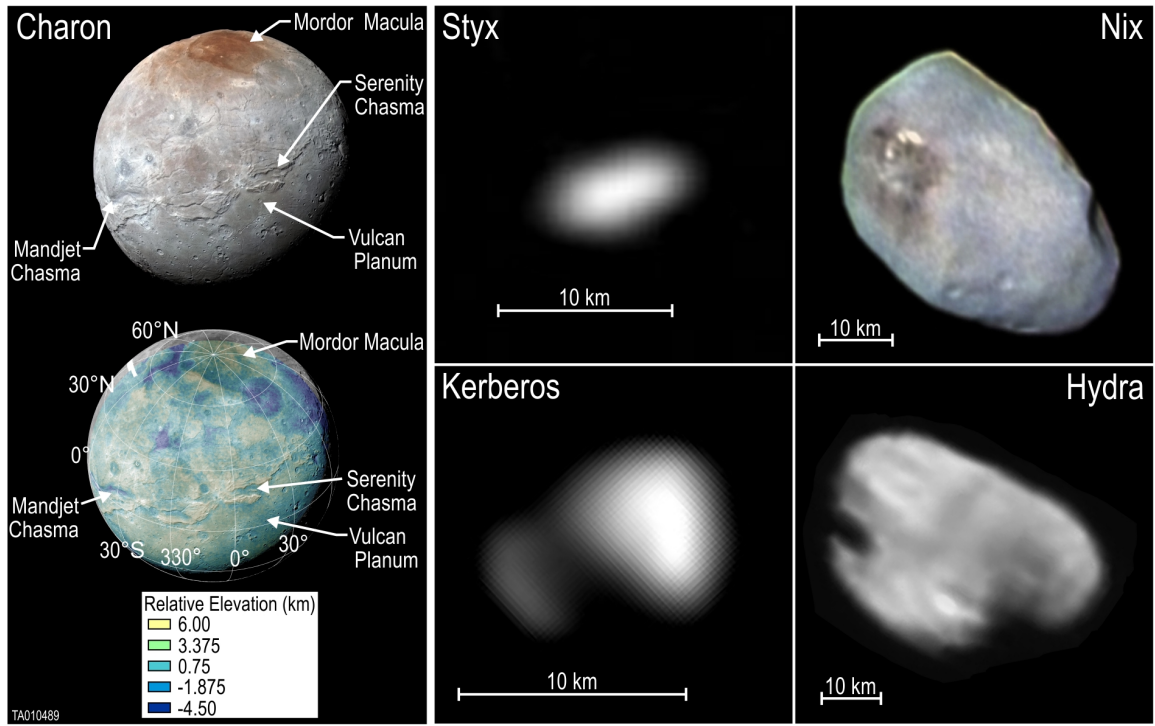

**Figure 8.** Enhanced NH color image of Pluto's largest moon Charon with a few major regions of interest identified by their informal names (adapted from Grundy et al. 2016a), and below that Digital terrain model of Charon showing the wide range of surface elevations. Adapted from Beyer et al. (2017). Also resolved images of Pluto's four small moons taken during the NH flyby. Celestial north is up and east is to the left. All images were deconvolved to recover the best available resolution. Panchromatic LORRI images were used for Styx, Kerberos, and Hydra, while an enhanced color image combining both MVIC and LORRI data was used for Nix. Some surface features on Nix and Hydra are impact craters. The largest crater on Nix is slightly darker and redder than the rest of Nix's surface. Adapted from Weaver et al. (2016).

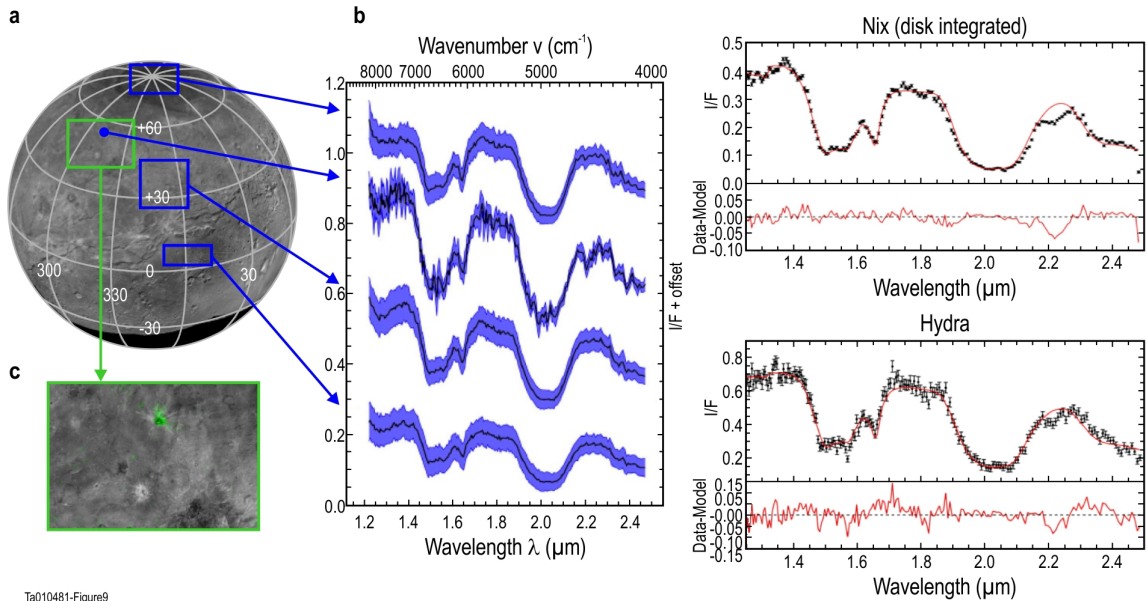

**Figure 9.** NH LEISA near-infrared spectra of Charon (left; adapted from Grundy et al. 2016a) and the small satellites Nix and Hydra (right; adapted from Cook et al. 2017a). All the spectra show deep absorption bands centered near 1.5, 1.65, and 2 μm associated with crystalline water ($H_2O$) ice. In the left figure, A is a LORRI composite base map. Regions where LEISA spectra were averaged for plotting in B are indicated by blue and green boxes. The different Charon spectra are generally similar, except "B" is a region near Organa crater showing additional absorption near 2.22 μm associated with ammonia-bearing species. An expansion of the region near Organa is displayed in C, and the region with the ammonia-bearing species is highlighted in green. The red curves in the figure on the right-hand side are from model fits that include only crystalline water ice. The plotted residuals show that both Nix and Hydra have an additional absorption feature near 2.2 μm that is attributed to ammonia-bearing species, and which may be associated with the same ammonia-bearing species seen in Organa crater on Charon.

**Table 1**

**Bulk properties of Pluto and Charon**[a]

| | Pluto | Charon |
|---|---|---|
| Radius (km) | 1188.3 ± 0.8 | 606.0 ± 0.5 |
| Mass (kg) | (1.303 ± 0.003) × $10^{22}$ | (1.586 ± 0.015) × $10^{21}$ |
| Density (kg m$^{-3}$) | 1854 ± 6 | 1702 ± 17 |
| Surface gravity (m s$^{-2}$) | 0.62 | 0.29 |
| Escape velocity (km s$^{-1}$) | 1.2 | 0.59 |
| Percent rock[b] by mass | 65.5 ± 0.5 | 59.0 ± 1.5 |
| Oblateness | <6% | <5% |

[a] All uncertainties are 1σ. [b] Calculated on an anhydrous basis.

**Table 2**
**Properties of Pluto's satellites**

| Object | Size[a] (km) | Orbital Distance[b] (km) | Orbital Period[b] (days) | Rotation Period[c] (days) | Rotation Pole[c] [RA,DEC] (deg) | Albedo[d] |
|---|---|---|---|---|---|---|
| Charon | 1212±1 | 19,573 ±2 | 6.387227 ±0.0000003 | 6.387227 ±0.0000003 | [132.993, -6.163] | 0.41±0.02 |
| Styx | 16x9x8 (10.5) | 42,656 ±78 | 20.16155 ±0.00027 | 3.24 ±0.07 | [196,61] | 0.65±0.07 |
| Nix | 48x33x30 (36) | 48,694 ±3 | 24.85463 ±0.00003 | 1.829 ±0.009 | [349,-38] | 0.56±0.05 |
| Kerberos | 19x10x9 (12) | 57,783 ±19 | 32.16756 ±0.00014 | 5.31 ±0.10 | [222,72] | 0.56±0.05 |
| Hydra | 50x36x32 (37) | 64,738 ±3 | 38.20177 ±0.00003 | 0.4295 ±0.0008 | [257,-24] | 0.83±0.08 |

[a]Size refers either to the diameter (Charon; Nimmo et al. 2017) or to the best fit tri-axial ellipsoid dimensions (small satellites; Porter et al. 2017). The size values in parentheses are the diameters for a sphere with the same volume as the ellipsoid. [b]The orbital parameters for Charon are from Buie et al. (2012), and the orbital parameters for the small satellites are from Showalter & Hamilton (2015). [c]The rotational parameters for Charon are from Buie et al. (2010a). Rotation Periods for the small satellites are from Weaver et al. (2016), and the Rotation Pole locations are updated values from Porter et al. (2017). Note the difference between the pole directions of the small satellites and that of Charon, the latter being perpendicular to the orbital plane of the system. [RA,DEC] refers to celestial coordinates in the standard J2000 system.[d] Albedo refers to the V-band geometric albedo. The values for the small satellites are from Weaver et al. (2016), and the value for Charon is derived